\documentclass[journal,10pt,twocolumn,twoside]{IEEEtran}

\usepackage[utf8x]{inputenc}
\usepackage{mathtools}
\usepackage{empheq}
\usepackage{algpseudocode, algorithm}
\usepackage{algorithmicx}
\usepackage{subfig}
\usepackage{varwidth}
\usepackage{url}
\usepackage{amssymb}
\usepackage{mathtools}
\usepackage{changes}
\usepackage[short]{optidef}
\usepackage{cite}
\usepackage{amsmath}
\usepackage{textcomp}
\usepackage{gensymb}
\usepackage{cite}
\usepackage{booktabs}
\usepackage{hyperref}
\usepackage{balance}
\usepackage{multirow}
\usepackage{makecell}

\begin{document}

\title{Blockchain for Edge of Things: \\Applications, Opportunities, and Challenges}

\author{Thippa Reddy Gadekallu, Quoc-Viet Pham, Dinh C. Nguyen, Praveen Kumar Reddy Maddikunta, \\N Deepa, Prabadevi B, Pubudu N. Pathirana, Jun Zhao, and Won-Joo Hwang


\thanks{Thippa Reddy Gadekallu, Praveen Kumar Reddy Maddikunta, N Deepa, Prabadevi B are with the School of Information Technology and Engineering, Vellore Institute of Technology, Tamilnadu, India (e-mail: \{thippareddy.g, praveenkumarreddy, deepa.rajesh, prabadevi.b\}@vit.ac.in).}

\IEEEcompsocitemizethanks{Quoc-Viet Pham (corresponding author) is with the Korean Southeast Center for the 4th Industrial Revolution Leader Education, Pusan National University, Busan 46241, Korea (e-mail: vietpq@pusan.ac.kr).}

\thanks{Dinh C. Nguyen and Pubudu N. Pathirana are with the School of Engineering, Deakin University, Waurn Ponds, VIC 3216, Australia, and also with the Data61, CSIRO, Docklands, Melbourne, Australia (e-mail: \{cdnguyen, pubudu.pathirana\}@deakin.edu.au).}


\thanks{Jun Zhao is with the School of Computer Science and Engineering, Nanyang Technological University, 50 Nanyang Avenue, 639798 Singapore (e-mail: junzhao@ntu.edu.sg).}

\thanks{Won-Joo Hwang is with the Department of Biomedical Convergence Engineering, Pusan National University, Yangsan 50612, Korea (e-mail: wjhwang@pusan.ac.kr).}

}




\markboth{IEEE Internet of Things Journal}{IEEE Internet of Things Journal}

\IEEEtitleabstractindextext{
\begin{abstract}
In recent years, blockchain networks have attracted significant attention in many research areas beyond cryptocurrency, one of them being the Edge of Things (EoT) that is enabled by the combination of edge computing and the Internet of Things (IoT). In this context, blockchain networks enabled with unique features such as decentralization, immutability, and traceability, have the potential to reshape and transform the conventional EoT systems with higher security levels. Particularly, the convergence of blockchain and EoT leads to a new paradigm, called \textit{BEoT} that has been regarded as a promising enabler for future services and applications. In this paper, we present a state-of-the-art review of recent developments in BEoT technology and discover its great opportunities in many application domains. We start our survey by providing an updated introduction to blockchain and EoT along with their recent advances. Subsequently, we discuss the use of BEoT in a wide range of industrial applications, from smart transportation, smart city, smart healthcare to smart home and smart grid. Security challenges in BEoT paradigm are also discussed and analyzed, with some key services such as access authentication, data privacy preservation, attack detection, and trust management. Finally, some key research challenges and future directions are also highlighted to instigate further research in this promising area. 
\end{abstract}

}

\maketitle
\IEEEdisplaynontitleabstractindextext
\IEEEpeerreviewmaketitle
\section{Introduction}
\label{Sec:Introduction}

In recent years, we have witnessed rapid advances in Internet of Things (IoT) empowered by the proliferation of mobile devices such as smartphones, laptops, sensors, wearables, etc. It is predicted that by 2030, the number of connected IoT devices surpasses 500 million \cite{d1}. This tremendous expansion of IoT is expected to create numerous applications and services across different application domains, from entertainment industry to mobile games and surveillance \cite{d2,d3,d4}. Such IoT applications often require high computing resources to handle massive data generated from sensor devices with latency requirements to provide time-sensitive customer services, like, transportation and smart healthcare. Cloud computing can support IoT devices in solving computation tasks, but high transmission latency remains a challenge due to long distance from the users. Edge computing has been recently proposed to support IoT with the creation of \emph{Edge of Things} (EoT) networks, by migrating computing and storage to the edge of the network, e.g., access points or base stations of radio access networks \cite{d5,d6,nkenyereye2021virtual}. In this regard, the computational burden posed on resource-constrained IoT sensors can be eliminated and the communication overhead is significantly reduced while providing better computing experience  for the users. Therefore, EoT possess the ability to support location-aware distributed IoT applications to facilitate time-sensitive service delivery with reduced computation complexities \cite{d7}.  

The distributed nature of EoT introduces new security and privacy challenges. The migration of large-scale computing and storage services to the edge creates the possibility of security threats and helps in controlling the network or prevent attacks on the resources at edge nodes \cite{d8,d9}. Moreover, uploading data to the network edge also raises critical data privacy concerns such as data breaches, data attacks and data modifications. Blockchain, a disruptive technology that emerged in recent years, has been regarded as a promising solution to solve security and privacy issues in edge computing networks as well as empower the next generations of EoT technology \cite{zhou2020solutions, wang2019survey,d12}. In particular, the convergence of blockchain and EoT creates a novel paradigm called \textit{BEoT}, which reshapes and transforms the conventional edge-IoT networks to enable new industrial and customer applications and services \cite{d10}, \cite{d11}. For example, BEoT has been used to provide secure smart city services such as reliable vehicular management and low-latency traffic control \cite{d20}. Further, BEoT has promoted smart healthcare analytics and IoT medical processing due to the large-scale computing and communication features of EoT and security features of blockchain \cite{d21}. The convergence of these emerging technologies is potentially a key enabler for future services and applications. 

\begin{figure*}
	\centering
	\includegraphics[width=1.0\linewidth]{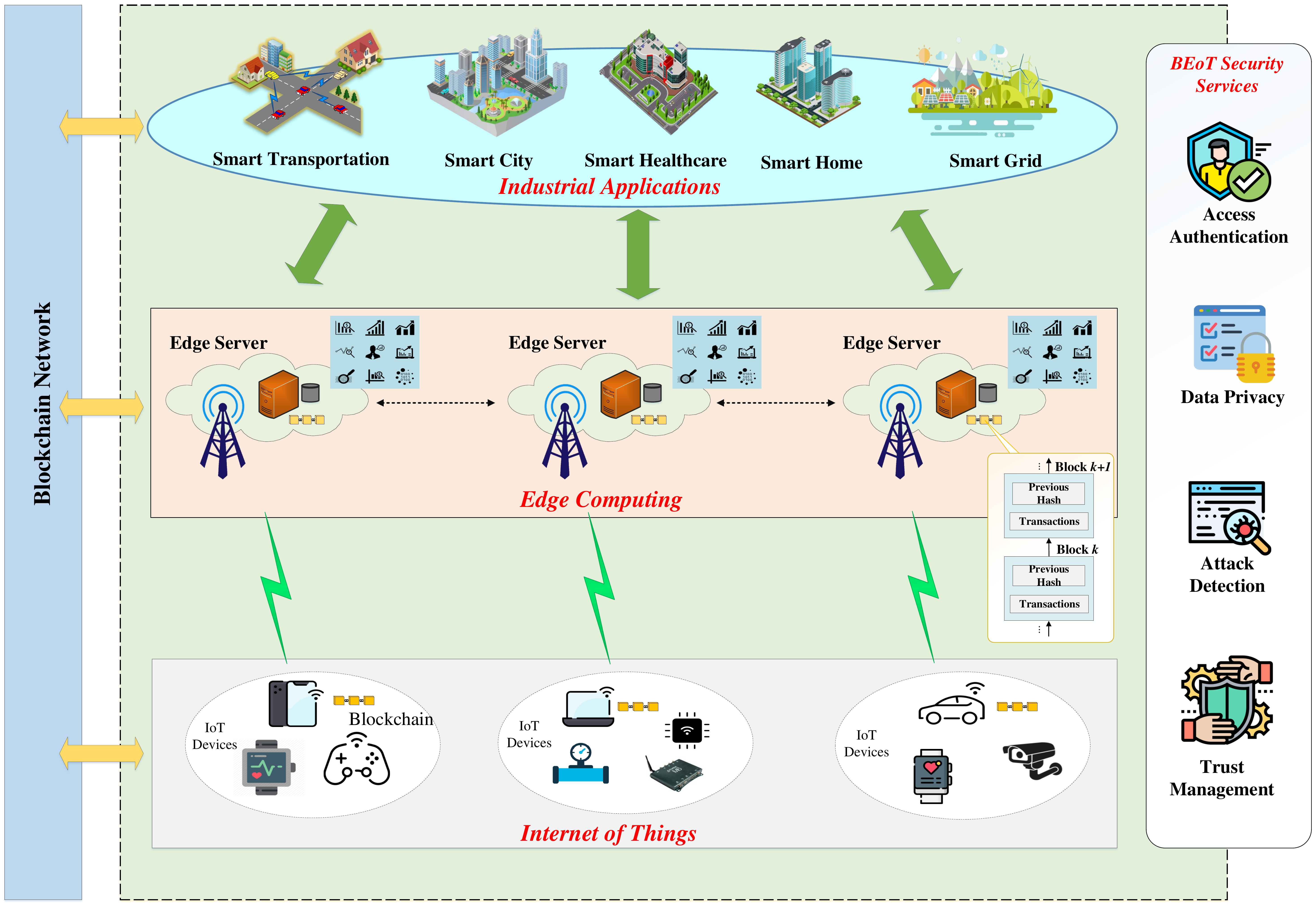}
	\caption{The generic architecture of BEoT.}
	\label{Fig:Overview}
\end{figure*}

\subsection{BEoT Architecture}
In this article, we propose a novel BEoT architecture that is enabled by the use of blockchain in EoT, as illustrated in Fig.~\ref{Fig:Overview}. The proposed architecture consists of three main entities; IoT, edge computing, blockchain, along with industrial applications and BEoT security services. 
\begin{itemize}
	\item \textit{IoT:} IoT devices such as sensors and mobile phones are responsible for generating or gathering data from the physical environments and then transmit to the nearby edge servers (ES) via access points or base stations. IoT devices with certain resources (e.g. smart phones, laptops) can act as a mobile blockchain entity to make transactions in order to communicate directly with the ES or even join the blockchain mining for extra profits \cite{d13}. Other lightweight IoT devices such as sensors can participate in the blockchain network via their representative gateways (e.g., mobile phones) or other mobile blockchain entities in its IoT network \cite{d14}. 
	\item \textit{Edge computing:} 
	In order to reduce the transmission time, it is necessary for computation nodes to perform data processing near to the end user. Due to heavy network traffic, cost of power consumption increases. To solve these issues, edge computing came into existence. It performs data storage and computing tasks in their edge network within short distance to the end user \cite{nkenyereye2021virtual}. As the edge computing nodes are closer to the users, the traffic flow is also reduced. It also minimizes the bandwidth demands and latency in data storage and computation in IoT network. In BEoT networks, IoT devices can offload their data to the ES located at the base stations for processing. ESs are typically equipped with rich computing and storage resources to handle IoT data tasks and provide data services for end users, ranging from data analytics, data prediction to data mining and data storage \cite{d15}, \cite{d16}. Moreover, each ES can also work as a blockchain miner that aims to verify the transactions and produce data blocks for maintaining the blockchain network. 
	\item \textit{Blockchain:} A blockchain is created to form the BEoT system running on top of the EoT network, aiming to interconnect IoT devices, ES and end users together in a decentralized fashion. Particularly, blockchain can guarantee the reliable operations of BEoT systems without the need of a central authority or third-party, by using some key services such as data consensus, smart contracts, and shared ledgers \cite{d18}. 
	\item \textit{Industrial applications:} BEoT enable new industrial applications, thanks to the application of blockchain in EoT. For example, in a BEoT-based smart transportation system, the secure data analytic services at the edge vehicular servers (i.e., roadside units) under the management of blockchain can support fast traffic control and reliable vehicle routing tasks even in the untrusted vehicular environments \cite{d19}. In the following sections, we provide a comprehensive discussion on the use of BEoT in various industrial applications, from smart transportation, smart city, smart healthcare to smart home and smart grid. 
	\item \textit{Security services:} Enabled by the inherent security properties such as decentralization, immutability, and traceability, blockchains provide a number of important security services for BEoT, including access authentication, data privacy, attack detection, and trust management. The analysis of such security services will be presented in detail in later sections. 
\end{itemize}
\subsection{State of the Arts and Our Contributions}
In the literature, many studies on blockchain and EoT topics have been investigated. The works in \cite{d22,d23,d24} present extensive surveys on the use of blockchain  and IoT from the different perspectives, spanning across various technical concepts, architectures to research challenges. The potential of blockchain networks in enabling IoT applications and services has also been investigated in \cite{d25}. Moreover, the possibility of combining blockchain and edge computing has been investigated and surveyed in \cite{d26}, \cite{d27}. The survey in \cite{d28} briefly discusses the role of blockchain in edge computing architectures. 

Despite so many research efforts, we have found that a comprehensive review of the use of blockchain in EoT is still missing. Moreover, reviewing the state-of-the-art in the field, BEoT has increasingly attracted much interest, both in academics and industry with a growing number of applied domains and use cases. Motivated by these, we provide an extensive survey on the applications of blockchain in EoT, ranging from applications, opportunities to research challenges and future directions in this article. The key objective of this article is to provide the readers with the state-of-the-art on blockchain and EoT and the recent advances in the BEoT technology. The key contributions of this survey are highlighted as follows.
\begin{enumerate}
	\item We provide a survey on the state of the art of the applications of blockchain in EoT networks, starting with an updated discussion on the recent developments of blockchain and EoT and highlighting the motivations of the use of blockchain in EoT. Moreover, a high-level BEoT architecture is also proposed and analysed. 
	\item 	 The key part of this article is focused on the opportunities of BEoT in industrial applications. In this regard, we present an in-depth survey on the use of BEoT in various important application domains, including smart transportation, smart city, smart healthcare, smart home, and smart grid. 
	\item Furthermore, the security requirements of the BEoT paradigm are also investigated. In particular, we analyze the benefits of blockchain in providing key security services for EoT, such as access authentication, data privacy preservation, attack detection, and trust management. 
	\item Based on the extensive survey on the BEoT, we identify the potential research challenges and highlight some important future directions in BEoT.
\end{enumerate}
\subsection{Structure of The Survey}
The remainder of the article is organized as follows. Section \ref{Sec:State-of-Art} describes the state-of-the-art in blockchain and EoT. The motivations of the use of blockchain in EoT are also highlighted. In Section~\ref{Sec:Application_Opportunities}, we survey and analyze the recent development of BEoT technology in a wide range of industrial applications, including smart transportation, smart city, smart healthcare, smart home, and smart grid. The security opportunities due to BEoT paradigm are also presented and discussed in Section \ref{Sec:Security_Opportunities} with some key services, such as access authentication, data privacy preservation, attack detection, and trust management. Section \ref{Sec:Challenges_Future-Directions} provides some research challenges and outlines some possible future directions in the BEoT. Finally, Section \ref{Sec:Conclusion} provides the concluding remarks on the core assertions of the paper. A list of key acronyms used throughout the paper is presented in Table~\ref{tab:ACRONYMS}.

\begin{table}[]
\centering
\caption{ACRONYMS}
\label{tab:ACRONYMS}
\begin{tabular}{ll}
5G    & 5th Generation                                      \\
ACL   & Access Control Lists                                \\
AI    & Artificial Intelligence                             \\
DDoS  & Distributed Denial-of-Service                       \\
DLT   & Distributed Ledger Technology                       \\
DRL   & Deep Reinforcement Learning                         \\
EDM   & Event-Driven Messages                               \\
EDoS  & Economic Denial of Sustainability                   \\
EHR   & Electronic Health Records                           \\
EoT   & Edge of Things                                      \\
ES    & Edge Server                                         \\
ICT   & Information and Communication Technology            \\
IEEE  & Institute of Electrical and Electronics Engineering \\
IETF  & Internet Engineering Task Force                     \\
IIoT  & Industrial IoT                                      \\
IoT   & Internet of Things                                  \\
ISO   & International Organization for Standardization      \\
ITU   & International Telecommunication Union               \\
MEC   & Multi-access Edge Computing                         \\
PoS   & Proof of Stake                                      \\
PoW   & Proof of Work                                       \\
SDN   & Software Defined   Network                          \\
SGN   & Smart Grid Network                                  \\
SHM   & Structural Health Monitoring                        \\
SSS   & Secret Sharing Scheme                               \\
SV    & Smart Vehicles                                      \\
UAV   & Unmanned Aerial Vehicle                             \\
V2G   & Vehicle to Grid                                     \\
VANET & Vehicular ad-hoc network                            \\
VEC   & Vehicular Edge Computing                        
\end{tabular}
\end{table}

\section{Blockchain and EoT: State of the Art}
\label{Sec:State-of-Art}
This section presents the background and recent developments of blockchain, EoT, and \textcolor{black}{highlights the motivations of the use of blockchain in EoT}. 
\subsection{Blockchain}
\begin{figure}[t]
	\centering
	\includegraphics[width=1.0\linewidth]{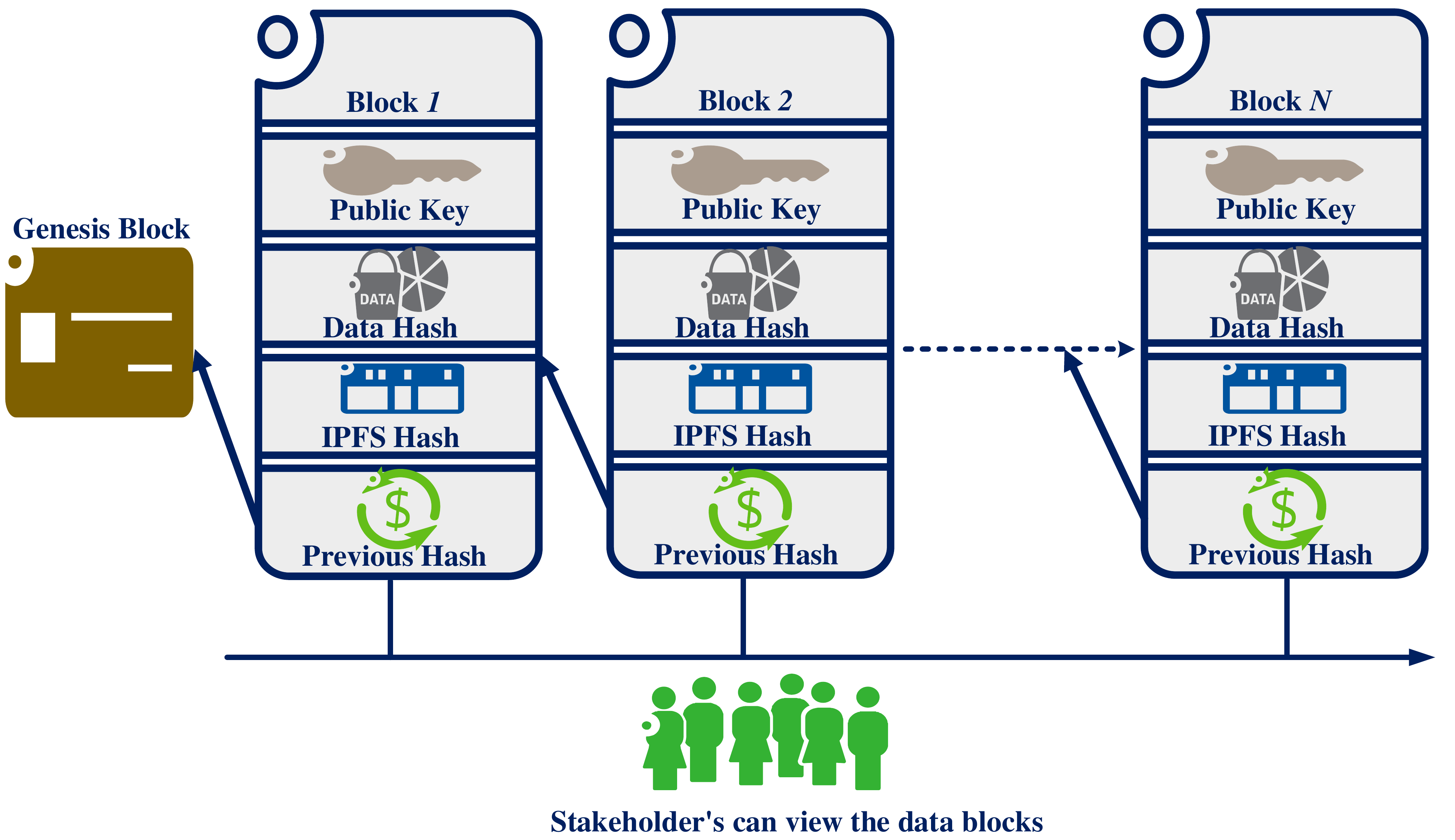}
	\caption{Blockchain Architecture.}
	\label{Fig:sec2.Block}
\end{figure}

Since the inception of Bitcoin, there is a huge buzz about blockchain. A blockchain is a chain of blocks, which is decentralized and distributed that can store information about transactions\cite{yaga2019blockchain}. \textcolor{black}{Each block on the blockchain is linked to its immediately-previous block through a hash label}. Specifically, a block in a blockchain can store the following information: (i) transaction details like time, date and value of transaction (ii) information about the person who is participating in the transactions and (iii) a unique hash code that differentiates a block from another block. For every transaction, a new block is created and added to the end of the blockchain. The blockchain is transparent as all the transactions in the blockchain are stored in a public blockchain and hence can be viewed by anyone. To maintain the privacy of the users, instead of storing their actual names or other information, the encrypted data will be stored in the public blockchain\cite{gai2019differential,baza2019b}. The block in the blockchain contains the hash of the information in it and also the hash of the block before it. Hence, blockchain is \textcolor{black}{often referred as} a decentralized distributed ledger technology\cite{feng2019survey,dwivedi2019decentralized}. The general architecture of a blockchain is depicted in Fig. \ref{Fig:sec2.Block}. 
If a hacker intends to edit the transaction in a blockchain, he has to modify hash of not only that block but also every other hash following it which is nearly impossible. This security property of blockchain makes it an ideal choice for usage in many sectors like banking, insurance, government services, supply chain management, etc.\cite{deepa2020survey}. 

Unlike traditional systems which require a central authority to verify and validate the transactions, the transactions in a blockchain are verified and validated by “consensus protocol”. A consensus algorithm is a mechanism where all the participating nodes in the blockchain network agree to the current state of the blockchain\cite{ kang2019toward}. Whenever a new block is created (by transactions) it requires a consensus algorithm to be executed so that all the nodes in the blockchain reach the consensus on the current state of the blockchain. The consensus algorithm is also executed when a new node is added to the blockchain. In this way, consensus algorithms ensure the reliability of the blockchain and also confirm each node in the blockchain as trustworthy\cite{ kumar2019proof}. Some of the popular consensus algorithms used today are Proof of Work (PoW), Practical Byzantine Fault Tolerance, Proof of Stake, Proof of Burn, Proof of Capacity\cite{xiao2020survey, ma2020efficient }. 

To ensure that the transactions meet the predefined terms and conditions, smart contracts are executed on a blockchain. A smart contract is a program that spans a few lines of code, that are used to make sure that all the transactions follow some kind of pre-agreements. Smart contracts ensure that the transactions are trustworthy. Smart contracts reduce the time which is otherwise spent on verifying the transactions. Accurate decisions can be made quickly because of verification of terms and conditions' verification is automated\cite{hakak2020securing}. The main reasons behind the popularity of blockchain are its unique properties, including decentralization, immutability and transparency.

\begin{itemize}
    \item \textbf{Decentralization}: Before blockchain came into limelight, a centralized entity used to store all the data and all the interactions with the data is through this centralized storage.  The centralized systems have several pitfalls like a single point of failure, vulnerability to attacks, etc. These drawbacks in centralized systems can be overcome by decentralized systems as every node in decentralized system stores the information.
    \item \textbf{Immutability}: Due to consensus algorithms, the information stored in the blockchain network is immutable. This property of blockchain makes it an ideal solution for usage in several sectors like finance, supply chain management sectors, governance, etc. \cite{politou2019blockchain}.
    \item \textbf{Transparency}: The technology used in the blockchain is always open-source. Even the transactions in the blockchain are transparent. The technology or the transactions are secured even though they are transparent as long as the majority of the blockchain network’s nodes have to approve the modifications. User information is hidden with the help of complex cryptography algorithms\cite{zhang2019blockchain}.
\end{itemize}
Even though there are many benefits, there are several key problems with the application of blockchain in distributed systems like IoT. A critical issue is the extensive energy consumption and high network latency caused by running consensus processes such as PoW in the blockchain. This may hinder the applications of blockchain in distributed EoT networks with resource-constrained IoT devices. Another problem is the limited throughput of blockchain systems. For example, Bitcoin can only process a maximum of four transactions/second, and the throughput of Ethereum achieved is about 20 transactions/second, while Visa can process up to 1667 transactions/second \cite{d29}. Moreover, security and privacy are other concerns to be considered when applying blockchain to EoT networks. For instance, a serious security bottleneck such as 51\% attack can prevent new transactions from gaining confirmations and halt payments between service providers and EoT users. Attackers can exploit this vulnerability to deploy attacks, such as transaction modifications, data breach, adversarial mining operations, all of which can degrade the performance of blockchain networks and results in data privacy leakage issues. Some solutions have been proposed to provide insights on solving these issues.  The work in \cite{ luu2017smartpool} provides lightweight consensus mechanisms to enhance the blockchain performance by compressing consensus storage and designing lightweight block validation schemes, aiming to simplify the blockchain mining process to achieve energy savings and latency improvement. Another study in \cite{ liu2019mathsf} introduces a mining pool system called SmartPool to improve transaction verification in blockchain mining to protect data privacy and mitigate security bottlenecks, such as 51\% vulnerability, ensuring that the ledger cannot be hacked by increasingly sophisticated attackers.
\subsection{Edge of Things}
With the improvement in communication technologies and affordable hardware, there has been a  rapid growth of smart devices in many areas of daily life and  \textcolor{black}{business} activities in  the past two decades. As information and communication technology (ICT) became affordable, there is an enormous surge in the data generated by mobile phones, IoT devices, industries. The volume of data generated resulted in the use of cloud for storage and computational purposes\cite{Pham2020ASurvey_MEC}. Storage and data processing in the cloud have their own challenges like latency, throughput, security, etc. For instance, storage and data processing in real time applications like traffic monitoring in the cloud may not be feasible due to increased latency \cite{Pham2020ASurvey_MEC}. The solution to this problem lies in edge computing, i.e., offloading the cloud-computing capabilities to the network edge \cite{khan2019edge,Pham2020ASurvey_MEC,mollah2017secure}. 

\begin{figure*}[t]
	\centering
	\includegraphics[width=0.825\linewidth]{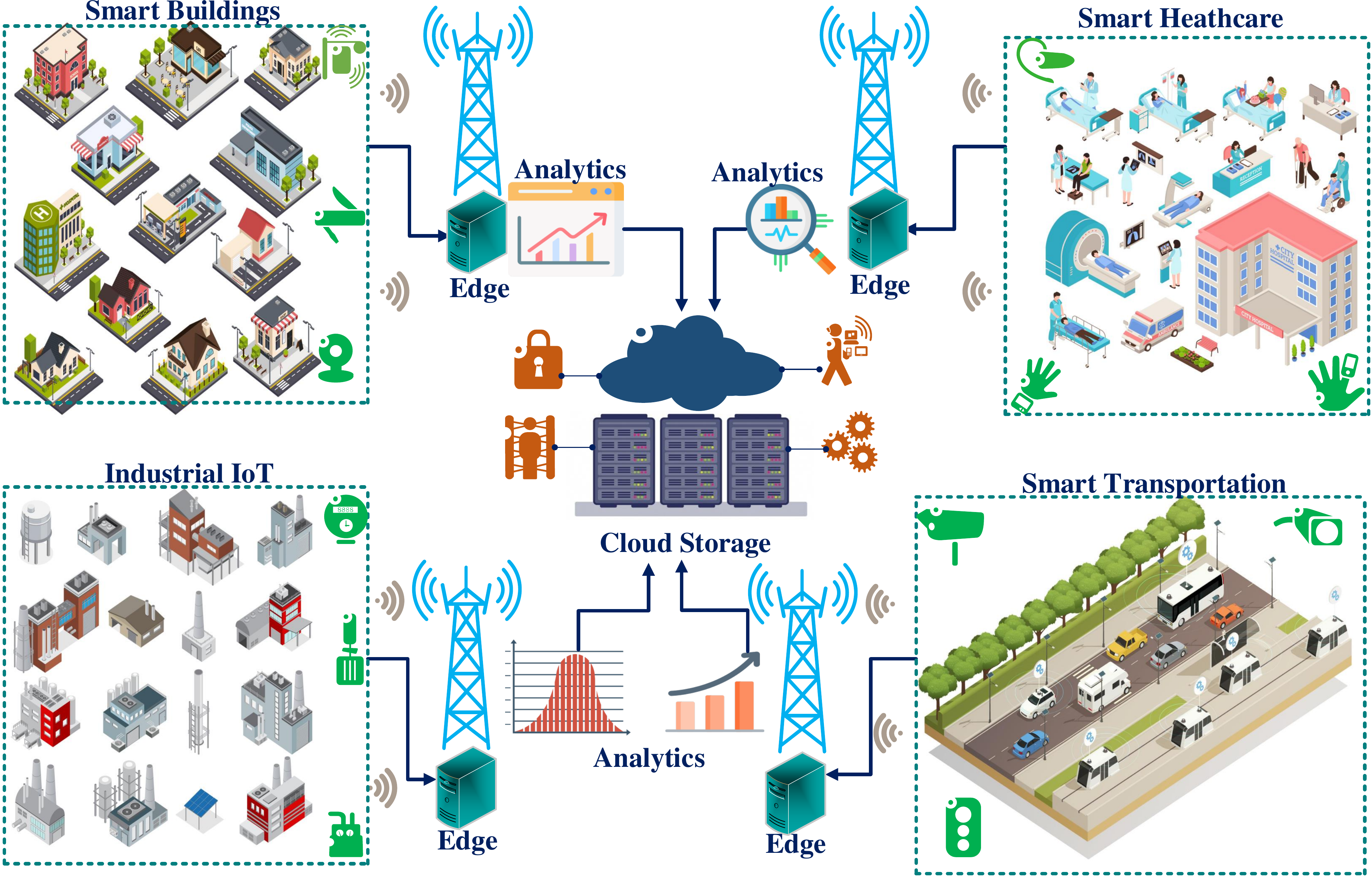}
	\caption{Architecture of EoT.}
	\label{Fig:sec2.1}
\end{figure*}

EoT is the integration of edge computing with IoT networks. In EoT, the data acquired by several sensors is temporarily stored in the edge node for real time analytics and predictions\cite{zhao2019computation}. The general architecture of EoT is depicted in Fig.~\ref{Fig:sec2.1}. Typically, the data generated from sensors in several IoT applications like smart homes/buildings, smart grids, smart healthcare, smart transportation, industrial IoT, etc. is stored in edge nodes at regular time intervals. Once the data is processed in the edge nodes it will be dissipated to the cloud. Apart from improved latency, EoT offers several other benefits like reduction of traffic to the cloud, improved reliability by installation of applications in close proximity to the edge devices, etc.

\subsection{Motivations of the use of Blockchain in EoT}

Even though multi-access edge computing is a promising solution for improved services of mobile providers, the security of the data in the edge nodes is a concern\cite{chen2020automated}. Several applications like connected vehicles, social media apps, healthcare related applications generate the data which is very sensitive\cite{alazab2020intelligent}. The privacy, confidentiality and integrity of these data \textcolor{black}{have to be} strictly maintained. The hackers can even attack the \textcolor{black}{Multi-access edge computing (MEC)} with several attacks like distributed denial-of-service (DDoS) attacks, hijacking of cloud servers, ripple attacks, byzantine attacks, injection attacks, etc. to steal the sensitive data from the edge or deny services to the users\cite{hussain2020deep}. 
Several smart city-based applications such as smart homes, smart grids that  generate sensitive data use edge nodes for real time analytics, \textcolor{black}{with a} fast throughput. Several industry 4.0 applications also use MEC for analytics in real time\cite{khan2020edge}. 
Several applications based on user location like Google Maps, \textcolor{black}{artificial intelligence (AI)} based Virtual Assistants use MEC for less latency and predictions/recommendations through machine learning \textcolor{black}{(ML)} based algorithms. All the applications mentioned above generate sensitive data of the users such as personal data, health, location, utility services, etc. may possess an elevated risk of compromised security. The properties of blockchain \textcolor{black}{such as} distributed nature, traceability, immutability make it an ideal solution to overcome the potential aforementioned problems by applications based on MEC. Blockchain has the \textcolor{black}{ability} to prevent issues like identity theft, DDoS attacks, tampering of transactions, user privacy leakages. \textcolor{black}{The use of blockchain in EoT} has the potential to be the next revolution in ICT where the mobile application providers can provide safe, transparent, immutable, decentralized applications to the customers with reduced latency, and real-time analytics/recommendations.

\textcolor{black}{The main motivations behind the use of blockchain in EoT are summarized as follows:}

\begin{itemize}
    \item The distributed architectures of EoT will provide a better roof for storing and verifying blockchain transactions.
    \item By using blockchain, data privacy and security can be well preserved in blockchain-enabled EoT applications. 
    \item The immutability and traceability features of blockchain can be leveraged to ensure the reliability of the transactions in industrial applications such as smart grid, smart transportation, smart health care, government services etc. 
    \item The consensus mechanism of the blockchain guarantees the trustworthiness and transparency of information transferred over the BEoT network.   
    \item The application of blockchain in EoT ensures low-latency response which is increasingly vital for most of the industrial applications.
    
\end{itemize}

\section{Industrial Applications from BEoT Paradigm }
\label{Sec:Application_Opportunities}
\begin{figure*}[t]
	\centering
	\includegraphics[width=0.925\linewidth]{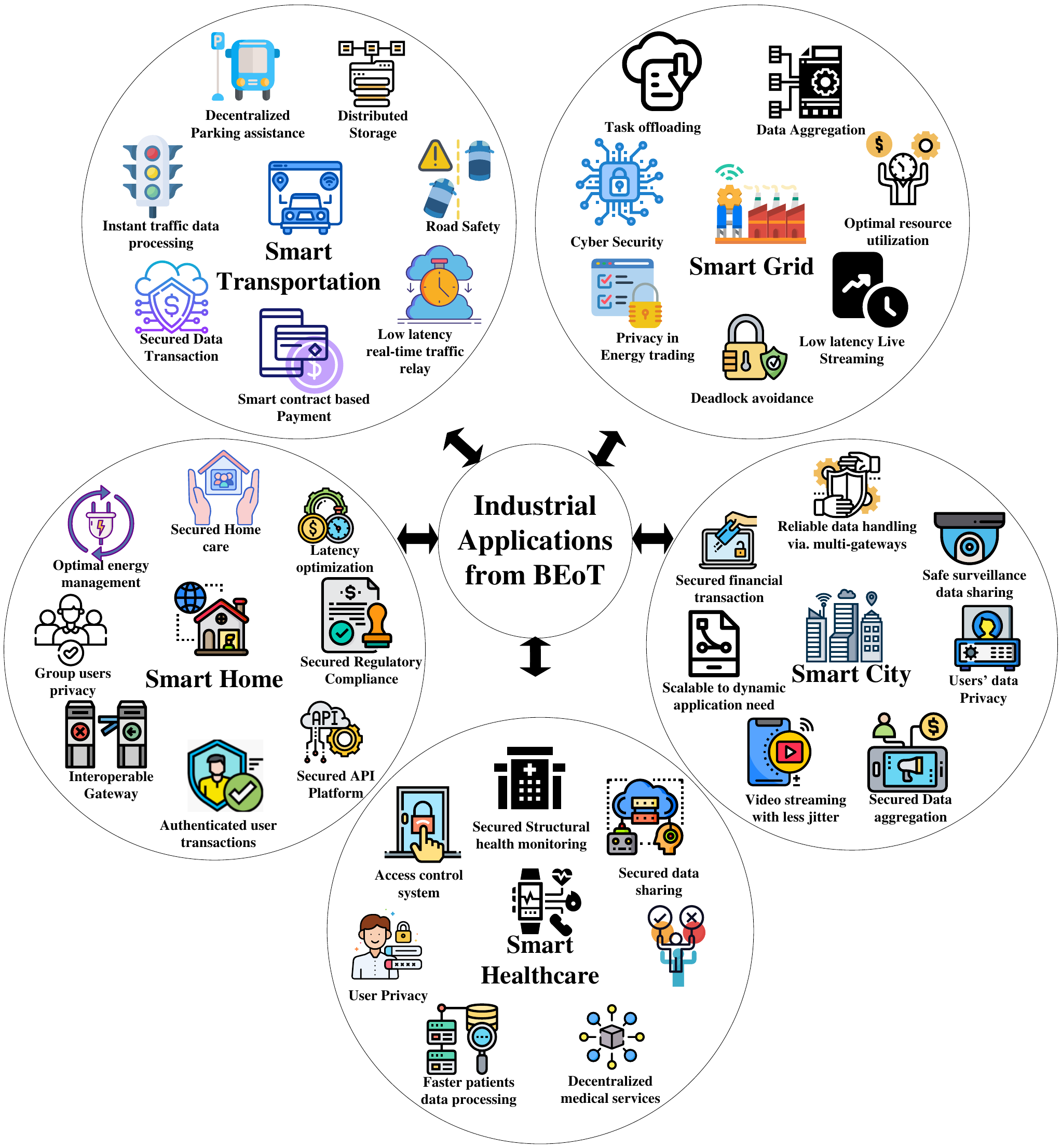}
	\caption{\textcolor{black}{Industrial applications of BEoT paradigm.}}
	\label{Fig:sec4}
\end{figure*}

This section presents the industrial applications {of the} BEoT paradigm {highlighting} some key benefits. Blockchain with EoT helps to modernize the large computer networks by providing smart architecture to various application domains such as  smart transportation, smart grid, smart city, smart healthcare and smart home. These benefits attained through BEoT in aforementioned industrial applications are depicted in Fig.~\ref{Fig:sec4}.

\subsection{Smart Transportation}
\textcolor{black}{The term \emph{smart} refers to the idea which helps to build an environment connected with sensors and other computing \textcolor{black}{facilities} for better understanding and controlling the user environment.}
Smart vehicles (SV) have garnered significant attention in recent times due to the advancement of ICT. Smart transportation system enables SVs to get connected to \textcolor{black} {the} Internet to access the required data and communicate with each other.  The aim of smart transportation system is to provide convenience and comfort to passengers and drivers. It also helps to improve the traffic efficiently and ensures road safety. Vehicles are connected to various network interfaces like WiMax, Bluetooth, and WiFi to communicate with road side units and other vehicles in smart transportation system \cite{xie2019survey,mollah2020blockchainiov}.

Smart transportation systems play a major role in the development of smart cities to keep track of traffic data and to avoid congestion, pollution, accidents etc. Due to the increase in the traffic data, the conventional centralized {approach} has faced many non-trivial challenges like storage of data, server failure, security, intelligent management etc. If the central server fails, the entire traffic system will be collapsed. Hence a decentralized solution is highly needed. The work in \cite{jiang2018blockchain} solved the problem by proposing a network model and {a} blockchain architecture.  In this model, {the} blockchain is integrated with vehicular networking application to provide security and distributed storage {for} \textcolor{black}{large amounts of data}. Several edge nodes are defined in the vehicle networks such as roadside vehicle and they form sub-blockchain networks. Data blocks in IoV are classified to form several blockchain networks in IoV system.

A platoon driving model was proposed for autonomous vehicles in an urban heavy traffic scenario to avoid congestion, accidents, and pollution. This model groups the vehicles by matching the path successfully in a platoon. Blockchain is integrated in this model in which smart contract is applied for payment purpose{s} which helps in overcoming false and malicious payment transactions. The results proved that the platoon model performed well for individual vehicle mode{ls} with respect to fuel consumption \cite{chen2019smart}. Parked vehicle assisted fog computing chain (PVFC) was introduced to accomplish decentralization using blockchain with smart contracts. Also\textcolor{black}{,} smart contract design was investigated to monitor the requester and performer behavior with high end privacy and security \cite{huang2020securing}. 

Smart transportation applications such as \textcolor{black}{self driving cars} produce large amounts of data of different types. To ensure safe driving, sharing of data is required to provide quality service while travelling. Due to lack of resources, vehicular network cannot provide huge data storage and data sharing. A distributed and secure vehicular blockchain was presented by exploiting blockchain consortium for the management of secure data in vehicular edge computing and networks \cite{kang2018blockchain}.

A multi-agent, autonomous, and intelligent management system was presented  in \cite{buzachis2020multi} for the safety of the vehicles passing through an intersection point. The system constructed using BEoT enables the communication between vehicle-to-infrastructure and infrastructure-to-vehicle. The system interacts with the vehicles in EoT environment and blockchain ensures the safety of pedestrians and drivers passing through the intersection point. The system helps to reduce the waiting and crossing time of the vehicles and ensures security and reliability. Blockchain supports reliability with decentralization and security by protecting the decisions taken from malicious attacks. \textcolor{black} {EoT helps in increasing the network performance,} thereby reducing the latency . 

\subsection{Smart Grid}
Electricity is one of the greatest inventions\textcolor{black}{,} without which today's digital advancements are impossible. Also, the electricity usage increases steadily  causing the production to increase. Traditional electrical grids use a centralized structure with millions of components such as power stations, substations, transmission lines and the distribution lines. It cannot accommodate new resource (increasing the load) as it may incur additional overhead leading to power quality issues, i.e. new plants have to be deployed whenever the load is increased. Also, the conventional grid doesn't have a proper prediction system on power blackout, slower response time, insufficient storage and  not efficient use of resources. Smart grid overlays the way for smart utilization of the electricity with fewer power outages and lower computational overheads \cite{gungor2011smart,alazab2020multidirectional,bashircomparative}. The smart grid comprises of smart meters assisted with mobile apps for real-time monitoring of power consumption, electric vehicles (EV), on-demand pricing capability, microgrid, storage, decision support systems, and other smart devices.  Smart power grids with its decentralized, distributed framework can strengthen the electrical power of a country through the effective utilization of renewable power resources and contemporary communication advancements. Consequently, this will reduce the power outages and ensures faster restoration of electricity after {blackouts}. 

Smart grid establishes two-way communication between the producer and consumer thereby focusing on making all the consumers as prosumers (they can simultaneously produce and consume electricity). Though smart grid is providing effective services, it has various challenges to be addressed. Some of them are protecting power grids from malicious attacks (integrated security), interoperability in connecting heterogeneous power systems, predicting the stability of smart grids, restoration, determining changing demand patterns, on-demand pricing, lack of regulatory policies, fault detection, and energy management \cite{zhang2017survey,pham2021deep}. To ensure secured transactions and to reduce the overhead{s} involved in data processing, blockchain and edge computing can be integrated with smart power grids. 

Security concerns of the smart grid network (SGN) include energy security and data security. 
EVs offer energy management solutions in SGNs through effective energy storage mechanisms. EVs are integrated with the power system and \textcolor{black}{store} the power from the grid and load it back to the grid whenever required. The energy trading (charging and loading the EVs) in the SDN-enabled vehicle-to-grid (V2G) infrastructure is a challenging issue. V2G technology of the smart grid reduces the level of demand-supply disparity by strengthening the energy trading capability of EVs. The SURVIVOR-Energy trading in SDN enabled V2G network using blockchain and edge computing framework presented in \cite{jindal2019survivor}, attempts to cover the tradeoffs in V2G environment. The energy trading decisions are taken at edge nodes closer to the EV to \textcolor{black}{reduce the effects of} the processing time, thereby reducing the latency and blockchain is employed to provide security in energy trading transactions. Though the results proved that blockchain is lightweight in terms of communicational and computation cost, content caching, and vehicle mobility remains to be addressed. Similar to \cite{jindal2019survivor}, a secure framework for energy trading between EV and grid (V2G) in cyber-physical systems was implemented in \cite{zhou2019secure}. Blockchain is used to secure the transactions in energy trading, followed by a contract theory based-incentive mechanism for V2G energy trading. The moderate cost consortium blockchain framework ensures secure V2G energy trading. An incentive contract theory based mechanism attains optimal resource utilization at LEPG. Edge computing is \textcolor{black}{utilized for task offloading with reduced latency and in turn, increases success rate probability in block creation, thereby reducing the computational overhead at local energy aggregators and assists in successful block creation.} Results show that there is a 124.6\% increase in the successful probability of block-creation. 

The convergence of blockchain, edge computing, cryptographic algorithm, and other techniques in the smart grid should not degrade the performance of the SGN. \textcolor{black} {In} \cite{wang2019blockchain}, a lightweight blockchain consortium was presented to ensure the scalability and efficiency of the SGN. Blockchain smart contract is used for managing the key materials table (public key identities) anonymously without exploiting the sensitive information. The security requirements for a smart grid with edge computing are detailed with Proof of Concept, and the model can combat known attacks such as replay attack, stolen verifier attack, and impersonation attack. The model proposed for key management ensures the efficient key update with less computational and communication overhead in comparison to similar models.
Henceforth, blockchain with data aggregation can be utilized in SGN for privacy-preserving transactions. Edge computing can be recommended for faster processing or task offloading, thereby providing a low-latency response. Also, to reduce the computational and communication cost with blockchain, a lightweight blockchain framework is recommended. 
\textcolor{black} {Various} industrial applications of blockchains integrated with EoT framework are shown in Table~\ref{tab:my-tab5}. It is evident from the literature that BEoT addresses the crucial requirements of the smart systems. Some of them are security, low-latency response, optimized resource utilization, reliability, scalability, and interoperability.  Of the various industrial applications, conferred smart city is the one that encompasses the services of other smart systems. For instance, the smart grid supplies the energy required for the operation of other intelligent systems as smart grid utilizes the services of smart transportation for energy trading through EVs. Smart transportation offers various reliable transport services to smart home (safe and comfort life) as well as smart healthcare (instant health services). Blockchain is used for handling all the transactions among the heterogeneous devices in the smart systems ensuring the interoperability and security. ES provides low-latency response in data processing by bypassing the frequent access to the cloud server for all data processing. Thus, the BEoT paradigm paves more significant support in various industrial applications by ensuring interoperable, secured and privacy-preserving data processing. 

\subsection{Smart City}
As the global population is increasing, it is challenging to meet diverse requirements of service provisions from different users. One of the innovative applications of IoT is the development of smart cities worldwide. Smart cities have the ability to control, monitor, track large volumes of data collected from various sensors installed in the city and provide essential services\cite{vinayakumar2020visualized}. 

Smart surveillance system integrated with IoT technology is one of the vital component of smart city. Some of the smart surveillance applications are face detection, motion detection, license plate detection, and threat detection. A smart surveillance system for smart city was proposed in \cite{nagothu2018microservice} using microservice architecture and blockchain. The conventional surveillance system is based on monolithic architecture which performs operations such as recording and monitoring whereas it lacks scalability and mostly relies on centralized architectures, which potentially raises security bottlenecks. The proposed microservice architecture decentralizes the operations from various distributed edge devices and proposed scalable solutions to smart surveillance systems.

An architecture was proposed with blockchain to support spatio-temporal smart contract solutions for sharing economy parameters in smart cities integrated with IoT environment. In this architecture, two entities can execute any number of secure transactions using cognitive systems without the need of third party while sharing economy related services using IoT framework for data processing. Cognitive system thinks like human beings and consists of \textcolor{black}{ML} algorithms for pattern recognition, natural language processing and data analytics. The cognitive engine is a part of blockchain technology which reads the available data resources from edge nodes and acquire knowledge for reliable decision making. In economy sharing services, transactions are automatic and managed by intelligent cognitive engine without the human intervention \cite{rahman2019blockchain}.
A homogeneous ecosystem was proposed, namely SmartME, in which several applications can be expanded to multinational range by enhancing a shared open ICT framework built for processing, sensing, storage of resources in the network. Several technologies such as cloud, fog, IoT,  edge computing, blockchain, \textcolor{black}{ML} are required to control the smart city ecosystem\cite{bruneo2019}. A secure, scalable, distributed network architecture was presented to enhance the strength of evolving blockchain and software defined network (SDN) technologies in smart cities. The architecture includes the features of distributed and centralized network architectures. Edge nodes serve as central server for specific infrastructure and records the credentials and access rules. The edge network helps to reduce the bandwidth of the network and obtain minimum latency. Argon2, a key derivative function based on PoW method was introduced in the proposed architecture in order to improve the security, privacy and abstain leaking information to attackers in distributed smart city network \cite{sharma2018blockchain}. 

\begin{figure*}[t]
	\centering
	\includegraphics[width=0.825\linewidth]{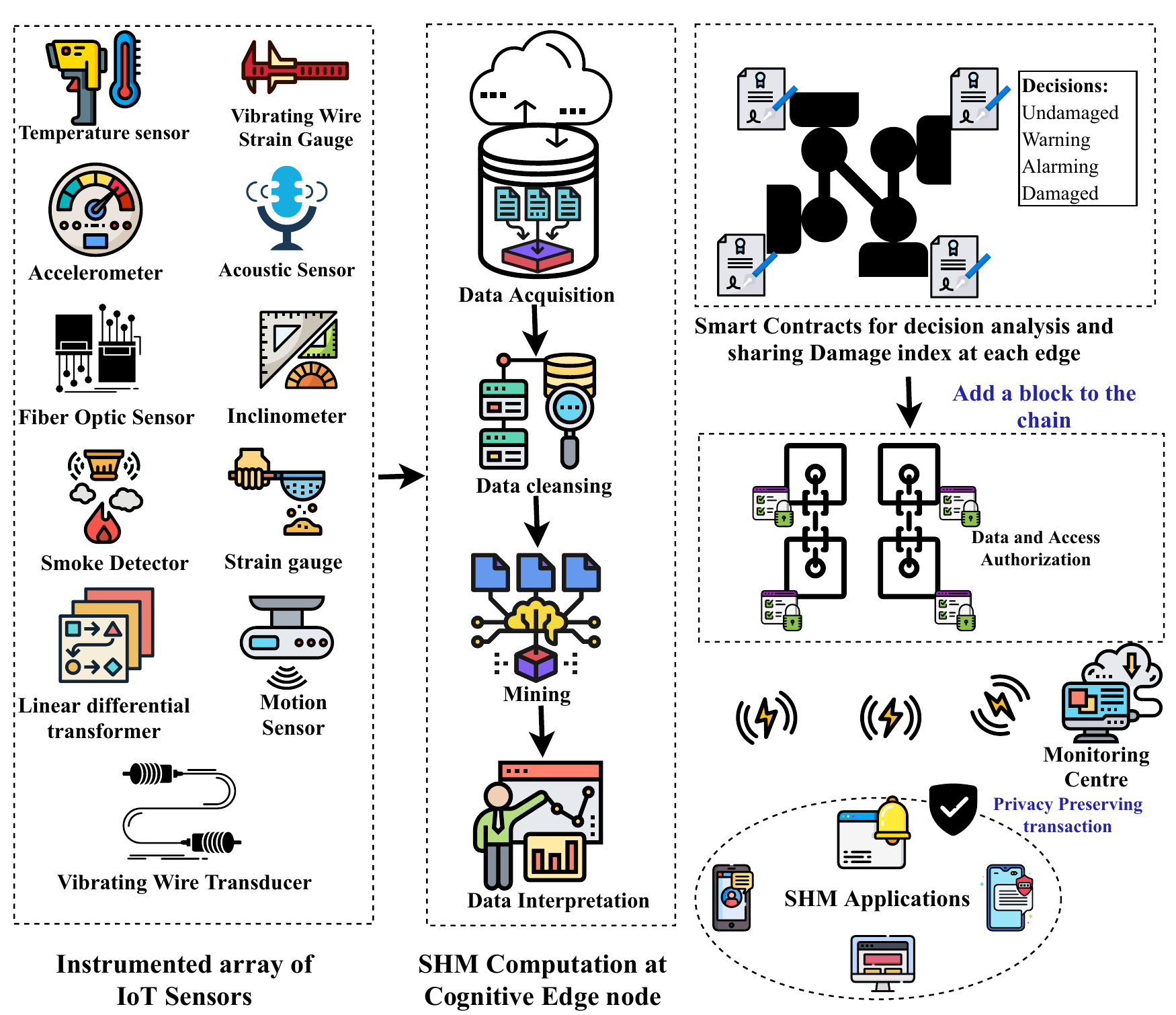}
	\caption{BEoT for secured structural health monitoring in smart city.}
	\label{Fig:BEoT_Healthcare_Smartcity}
\end{figure*}

A framework based on blockchain was presented in \cite{makhdoom2020privysharing} for IoT data sharing and privacy preserving in smart city applications. In this work, the blockchain network is divided into various channels and each channel processes the different type of data obtained from various edge nodes such as finance, smart car, smart energy, smart healthcare, etc. Each channel consists of number of certified organizations. Access control within the channel is managed by smart contracts. Security of data in each channel is achieved using encryption algorithms. Thus, BEoT has provided numerous solutions for the decentralization, security and interoperability problems in smart city based applications.
Another study was conducted in \cite{jo2018hybrid} by combining blockchain with IoT in structural health monitoring (SHM) systems, aiming to improve the operational safety for underground environments in smart cities. In this blockchain based IoT network, the centralized and decentralized distributions are provided by splitting the network into core and edge networks. These networks provide autonomous monitoring and control which improves the scalability and efficiency of the system. Also blockchain based decentralized networks can be deployed to provide efficient and transparent information sharing, security and decision making using smart contracts in SHM. 

A BEoT use case on secured SHM in smart city is presented in Fig.~\ref{Fig:BEoT_Healthcare_Smartcity}. The damages in the building are detected based on several factors. These factors are determined by data acquired from an instrumented array of heterogeneous IoT devices such as fibre optic sensors, temperature sensor, inclinometer (tiltmeter and slope detector), accelerometer, strain gauge, vibration sensor, acoustic sensor, smoke detector, transducer, and linear differential transformer.  The data is assimilated in the cloud server can be retrieved using the cognitive edge nodes. Cognitive edge nodes are intelligent enough to perform the mining of the data accumulated and early prediction of any problems in the structural health of the building. Smart contracts are used for making decisions, analysis of the type of problems and sharing the damage index of the miner among the edge nodes. For each transaction, a block is added in the distributed ledger assuring authorized access to data and the access privileges to the user. In turn, the notification on the structural health of the building (decisions made through smart contracts) is communicated to various applications and monitoring centres.
 
\begin{figure*}[t]
	\centering
	\includegraphics[width=0.975\linewidth]{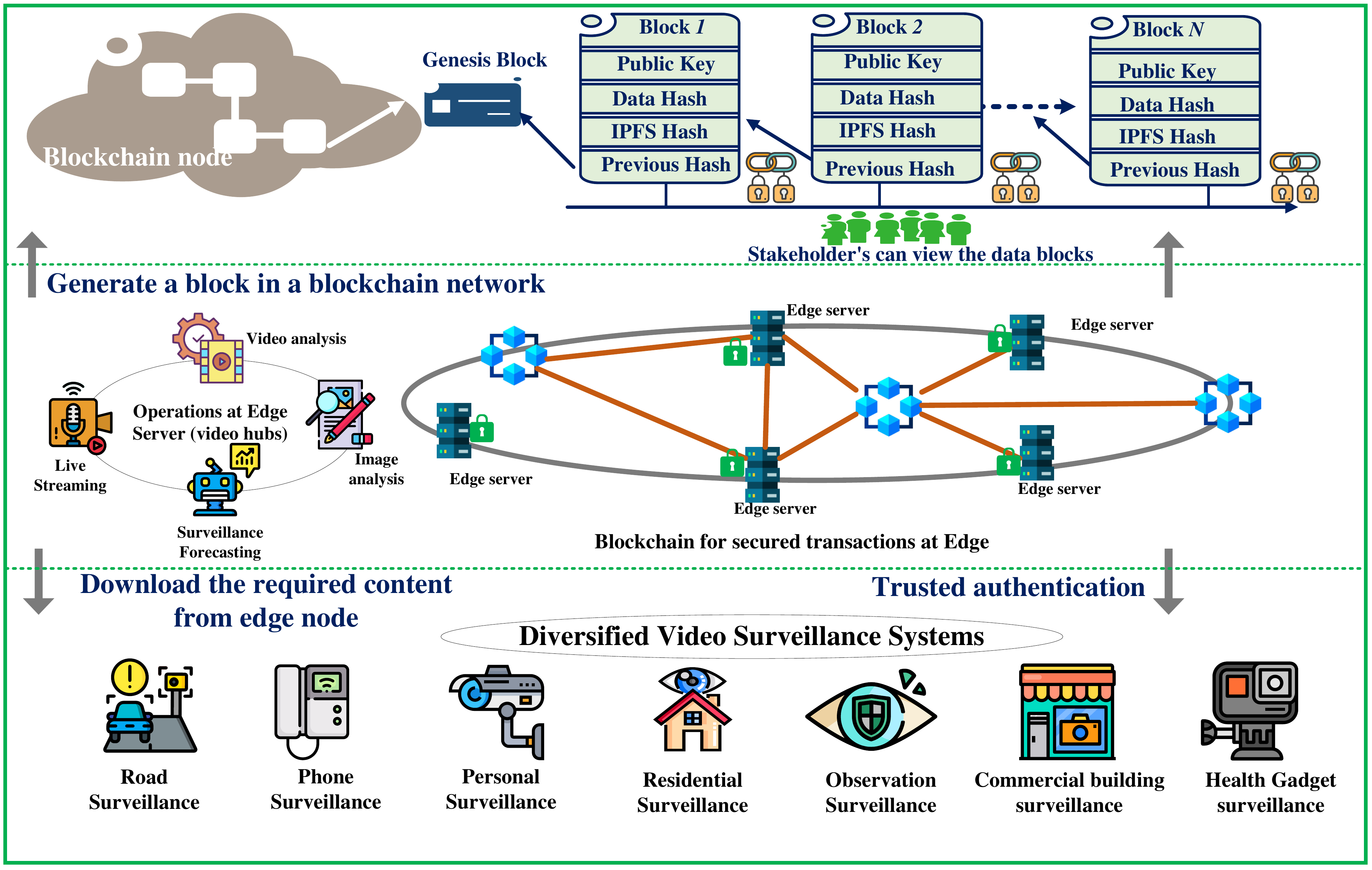}
	\caption{BEoT for safe surveillance data sharing in smart city.}
	\label{Fig:BEoT_safesurveillance_datasharing}
\end{figure*}

Another BEoT use case on safe surveillance data sharing is presented in Fig.~\ref{Fig:BEoT_safesurveillance_datasharing}. Most of the decisions in the smart city environment rely on the surveillance data. Any data transaction, involving these surveillance systems, should be done in a very secured manner. There are diversified video surveillance systems exists in the smart city environment such as video enabled telephone, private video surveillance, commercial building video surveillance, residential surveillance, observation surveillance, road traffic surveillance and health gadget surveillance. Real-time information from many these surveillance systems is required for taking many important decisions. For instance, road surveillance video helps to determine the traffic signal. In turn, crowd gathering information obtained from the commercial building video surveillance determines the reason for the traffic. Also, the acute health issues can be evaluated through health gadget surveillance and more. Videos from these diversified surveillance systems are  processed, analyzed and predicted by the video hubs on the edge nodes where blockchain-based smart contracts are deployed for secured and trusted authentication. Furthermore, to avoid overheads incurred in processing enormous data stored at blocks in the blockchain, side chains can be used to \textcolor{black}{segregate} unwanted information being stored in the blocks. This will enhance the efficiency of the overall system.

\subsection{Smart Healthcare}
Healthcare sector has succeeded \textcolor{black}{in being one of the leading} domains with respect to employment and income generation \cite{deepa2020ai,gadekallu2020deep}. Even though IoT is contributing to various domains such as smart home, smart agriculture and smart city, its impact on healthcare sector is remarkable. With the development of technologies like IoT, \textcolor{black}{AI}, 5th Generation (5G) network, mobile Internet, big data and cloud computing, traditional medical systems are transformed into smart healthcare systems. Using these advanced technologies, smart healthcare systems reduce the threat and cost of medical practices and enhances the progress of telemedicine. Smart healthcare is a medical service that applies technologies such as IoT, AI, wearable devices and mobile Internet to acquire information, link people, tools and organizations associated to healthcare, actively monitors and responds to medical assistance engaging intelligent system. Patients can monitor their health status by using wearable devices, get the medical services through virtual online support systems and the doctors can predict the diseases. Various problems such as security, privacy, transparency, interoperability, decentralization, data storage, and sharing must be addressed for the deployment of smart healthcare systems. Blockchain provides better solution to address these issues in smart healthcare and various research initiatives have been started by integrating blockchain with edge computing \cite{gupta2019habits}.

A secure healthcare scheme namely \textit{BHealth} based on blockchain was presented using unmanned aerial vehicle (UAV) in IoT. The UAV collects medical data from the users and stores in nearest MEC server. Basic user information is securely stored in a blockchain using smart contracts. Blockchain synchronizes the health data, secures the data with encryption, verifies the users and allows the UAV to store the data in the server\cite{islam2020blockchain,ch2020security}.

Blockchain based secure therapy management framework was presented in~\cite{rahman2018blockchain} using MEC for with disabilities in various age groups. The framework was developed by leveraging IoT nodes, blockchain and MEC. Blockchain with MEC provides decentralization, security, low-latency response and data sharing facilities for therapeutic data. The patient is allowed to share their data related to therapy with anyone. The mobile edge network processes the therapy data and prevents the limitations caused by high bandwidth. The related multimedia data such as audio, video and images are stored in a centralized or distributed storage based on the application.

Providing security and privacy to the patient data collected from various sources such as sensors, wearable devices etc., \textcolor{black}{stored} in electronic health records (EHR) database is another pivotal challenge in smart healthcare systems. A hybrid architecture using blockchain and edge computing was proposed to provide access control for the EHR data \cite{guo2019access}. All the access events are verified and stored using a mutual agreement mechanism before they are included to the blockchain. The EHR data is stored in edge nodes, which apply access control policies to provide attribute based access in association with blockchain. The access control list is executed by the healthcare providers to obtain one time self destructing \textit{URLs} which contain the address of the EHR data. Then EHR data is accessed by the providers using the URLs. Hence, only authorized users who have access to attribute based access control provided by edge nodes can access the EHR data.

A spatial and secure blockchain based mass screening framework  was presented for data storage, sharing and management of dyslexia data. The framework analyzes the data and predicts the symptoms of dyslexia. A mobile based multimedia IoT environment was presented to capture the user interaction of dyslexia testing data from a smartphone and share it in the mobile edge network. The edge node applies auto grading algorithms on the data for predicting dyslexia symptoms and the final results are stored in the blockchain. Blockchain provides decentralized data repository for captured multimedia based IoT test data shared for medical research and analysis \cite{rahman2018spatial}.

\subsection{Smart Home}
In this world of digital computation, people are committed towards their work, so it is harder for an individual to be constantly vigilant on household chores. Smart home, with ICT assisted devices make the home a safer and better place to live. One of the critical use cases of a smart city is a smart home. The smart home uses the Internet, to interconnect all the household appliances for facilitating seamless communication between the residents and the home appliances. Some of the remotely controlled functions in a smart home are closed circuit television, air conditioners, television, lighting systems, speaker, thermostat, temperature, refrigerator, doors, and pet feed. Any appliance with the capability of the remote access can join the smart home network and can be controlled through a laptop, PC, a smartphone or a tablet. This interconnection of objects is possible through IoT devices which makes use of the Internet to connect all the objects with the ability to share electronic information. The general types of alerts raised by these smart household devices are motion detection in case of theft, automatic control of home appliances upon its unwanted usage, healthcare of home ridden older people,  kids activity monitoring, etc. \textcolor{black}{The alert system} will send alert notification not only to the home residents but also to the concerned security providers in case of theft, the fire station in case of fire, blue cross services in case of animal ill-treatment and ambulance service for health care issues. Therefore the primary concern for implementation of smart home system is security and privacy. Also, this concern paves the way for a lot of security threats in terms of data theft and cyber-attack, such as DDoS exploiting all the network bandwidth for hijacking the environment. Furthermore, application of blockchain in the smart home helps to protect all the IoT devices in smart homes and data acquired from these devices for establishing interconnection among heterogenous applications while participating as an IoT node in the smart city blockchain network. 

Data access from the Internet-enabled smart home appliances will be easier for the adept hackers, the users of technology. So, the work in \cite{lee2020blockchain} presented a blockchain-based gateway architecture to prevent data theft by malicious users. Multiple security interventions in centralized gateway architecture of the smart homes are addressed, and severe network attacks are counterfeited. The data transaction among various devices in the smart home network are carried out only for the devices registered in the gateway. The blockchain is incorporated in this gateway layer of a smart home network which records and authenticates the devices joining the network through a SHA2 encryption algorithm, thus avoiding the data theft. The additional computational complexity incurred by blockchain at the gateway can be reduced by adopting edge computing for offloading the computation overhead. The proposed BEoT framework prevents the gateway from the attacks, namely blockchain 51\% attack, patch file forgery attack and DDoS attack.

{\begin{table*}[t]
\centering
\caption{\textcolor{black}{Unique requirements of blockchain in industrial applications.}}
\label{tab4}
\begin{tabular}{|p{0.5 cm}|p{1.5 cm}|p{4.5 cm}|p{4.7cm}|p{4.8 cm}|}
\hline
\textbf{S.No} &
  \textbf{Industrial applications} &
  \textbf{Common problem} &
  \textbf{Unique requirements of blockchain} &
  \textbf{Challenges in blockchain adoption} \\ \hline
1 &
  Smart Transportation &
  Stabilized network, uninterrupted and secured data   sharing among edge devices and other public vehicles  &
  Blockchain can be utilized for uninterrupted and   secured data sharing and privacy to retain the efficiency of ITS &
  Vulnerable to cybersecurity attacks during real-time   data sharing and optimized resource utilization \\ \hline
2 &
  Smart Grid &
  Secured energy trading among prosumers &
  Blockchain for secured and privacy-preserving   energy trading transaction in edge computing-based smart grids &
  Resource optimization (communicational and computaional resources) \\ \hline
3 &
  Smart City &
  Secure data sharing by users for facilitating varied   smart applications  (multiple trust   domains) &
  Blockchain for secured and transparent data-sharing   among users and multiple trust domains &
  Scalability issues when the number of transactions   increased \\ \hline
4 &
  Smart Healthcare &
  Distributed authorization of edge nodes, security and   data privacy in healthcare data records &
  Blockchain ensures data privacy in more sensitive   health care data transactions and restricted third party data access &
  Security issues pertaining to key management;   Scalability issues when the number of transactions increases and lack of   standardization \\ \hline
5 &
  Smart home &
  Access control issues to different residents and   visitors &
  Blockchain can be used for users privacy requirements   and enabling secure data sharing among non-interoperable that party service   providers &
  Vulnerable to security attacks while intact with the consensus process and scalability issues when the   number of applications increase \\ \hline
\end{tabular}%
\end{table*}
}

A secure authentication system integrating blockchain (for ensuring reliability in user transactions), group signatures (GS) (to authenticate different devices in the network) and message authentication code (for authenticating gateways) was implemented in  \cite{lin2019homechain}. This model not only provides a solution for critical challenges in blockchain applications, but it can also efficiently trace the footprints of the intruder misbehaviour.

Remote monitoring of patients in a smart home using a fog assisted IoT based in-house patient monitoring system was presented in \cite{verma2018fog}. To avoid the unprecedented delay caused by processing the data to and from the cloud, the three-layered model proposes the fog computing services with notification mechanisms at the network edge (i.e., in the gateway). The proposed model offers real-time interactive services on event classification with minimum latency at the fog layer. Fog assisted IoT model can effectively monitor various behaviors of the patients and provides real-time notification on the behavior of the patients with minimum delay in processing. The model performs the accurate classification of events based on the behavior of the patients in the smart home using BBN with temporal mining. 

Furthermore, a ChainSDI (Software Defined Infrastructure) framework implemented in \cite{li2019chainsdi} influences blockchain along with edge computing to provide a secure sharing and computation of smart home patients data. The framework attempts to address the data interoperability and regulatory issues in emerging SDIs used for healthcare applications. ChainSDI is an API on SDI that serves as a testbed for any healthcare application. Though ChainSDI provides better security and  privacy in handling users transactions, the communication and computation cost is increased. 

The industrial applications discussed in this section used edge computing enabled services for low latency response. Also, these industrial applications interact with each and exchange their services. The edge enabled smart environment faces significant challenges {in} security and privacy {space} such as authentication, access control, intrusion detection as the ES are heterogeneous and migration of services across these servers are prone to various security threats. Blockchain with EoT can address these issues, and the unique requirements of blockchain in these smart applications are depicted in Table~\ref{tab4}.

\begin{table*}[h!]
\centering
\caption{Survey of industrial applications of BEoT paradigm.}
\label{tab:my-tab5}
\resizebox{\textwidth}{!}{%
\begin{tabular}{|c|c|p{2.9 cm}|p{2.9 cm}|p{9cm}|}
\hline
\textbf{Applications} &
  \textbf{Ref.} &
  \textbf{Contribution} &
  \textbf{Technologies used} &
  \textbf{Key features} \\ \hline
  \multirow{8}{*}{\makecell{Smart \\transportation}} &
  \multirow{2}{*}{\cite{chen2019smart}} &
  Platoon driving model for urban IoVs &
  Blockchain, IoV, edge cloud &
  A vehicle platooning mechanism assists to obtain path information matching, smart contract based payment mechanism in urban road traffic condition \\ \cline{2-5} 
  & \multirow{2}{*}{\cite{huang2020securing}} &
  Parked Vehicle assisted fog computing chain &
  Blockchain, smart contract, fog computing &
  Provides decentralization and security for parked vehicle assistance in vehicular network using blockchain with smart contracts \\ \cline{2-5} 
  & \multirow{2}{*}{\cite{kang2018blockchain}} 
  & \multirow{2}{*}{Vehicular blockchain} 
  & Blockchain and edge computing &
  Utilizes vehicular blockchain and smart contracts to obtain   data storage, sharing and security in vehicular network \\ \cline{2-5} 
  & \multirow{2}{*}{\cite{buzachis2020multi}} &
  Multi-agent road safety system &
  \multirow{2}{*}{Blockchain, EoT} &
Ensures safety and  security using blockchain, and enhances network performance and latency reduction using EoT \\ \hline
 \multirow{6}{*}{\makecell{Smart \\ \textcolor{black}{grid}}} &
  \multirow{2}{*}{\textcolor{black}{\cite{jindal2019survivor}}} &
 \textcolor{black}{Energy trading in SDN enabled V2G network} &
 \textcolor{black}{SDN,Blockchain,Edge computing and EVs}
&
\textcolor{black}{SDN enabled EVs offer less latency and Lightweight-blockchain with reduced computational overhead provides security in processing the energy transaction.}
\\ \cline{2-5} 
  & \multirow{2}{*}{\textcolor{black}{\cite{zhou2019secure}}} &
  \textcolor{black}{Secure V2G energy trading} &
  \textcolor{black}{Blockchain, Edge and contract theory}
 &
\textcolor{black}{Blockchain ensures secure V2G energy trading, contract theory optimal resource utilization and EC ensures task offloading with low latency}
 \\ \cline{2-5} 
  & \multirow{2}{*}{\textcolor{black}{\cite{wang2019blockchain}}} 
  & \textcolor{black}{{Mutual authentication system in SGN}} 
  & \textcolor{black}{Blockchain and edge computing} &
  \textcolor{black}{Almost all the security requirements of the edge enabled SGN was met with lower computation and communication costs for key management} 
\\ \hline
    \multirow{8}{*}{\makecell{Smart \\city}} &
 \multirow{2}{*}{\cite{   rahman2019blockchain}} &
  Sharing   economy services in smart city &
  BIoT, cognitive edge nodes assisted with AI &
  Financial transactions are   automatic and managed by intelligent cognitive engine in Blockchain without   the involvement of human using edge computing \\ \cline{2-5} 
 &
  \multirow{2}{*}{\cite{bruneo2019}} &
   \multirow{2}{*}{SmartME} &
  Fog, BEoT and ML &
  SmartME scales up the   applications to wide range by enhancing open sharable ICT and applies edge,   fog, blockchain to control the smart city ecosystem \\ \cline{2-5} 
 &
  \multirow{2}{*}{\cite{   sharma2018blockchain}} &
  Hybrid   network architectural framework &
  \multirow{2}{*}{BEoT and SDN} &
 Offers the features of both distributed and  centralized architectures. Edge node serve as central server and  records the credentials thereby reducing the latency \\ \cline{2-5} 
 &
  \multirow{2}{*}{\cite{   makhdoom2020privysharing}} &
  Secure   framework for IoT data sharing &
  Blockchain,   IoT, edge computing &
  The framework divides the network into multiple channels and each channel secures the data related to specific application collected from edge devices   \\ \hline
\multirow{8}{*}{\makecell{Smart \\healthcare}} &
    \multirow{2}{*}{\cite{   islam2020blockchain}} &
   \multirow{2}{*}{BHealth} &
  Blockchain, smart contract and MEC &
  The scheme synchronizes the health data,  secures   the data with encryption, verifies the users and allows the UAV to store the data in the ES \\ \cline{2-5} 
 &
  \multirow{2}{*}{\cite{   rahman2018blockchain}} &
  Therapy management framework &
   \multirow{2}{*}{BIoT and MEC} &
  The framework for the differently abled people provide decentralized,  secured, low-latency response and therapeutic data sharing facilities \\ \cline{2-5} 
 &
  \multirow{2}{*}{\cite{   rahman2018spatial}} &
  Blockchain based mass screening framework &
  BEoT in mobile and auto-grading algorithms &
Provides decentralized data repository for captured multimedia based IoT test data   shared for medical research and analysis. \\ \hline
  
  \multirow{2}{*}{\makecell{Smart \\ home}}  &
  \multirow{2}{*}{ \cite{ li2019chainsdi}} &
  ChainSDI, regulatory compliance &
  SDI,   Edge  Computing and Blockchain &
  Provides secured specification for regulatory compliant requirement in data processing and a low-latency response in health care-related data processing \\ \hline
\end{tabular}%
}
\end{table*}

\section{\textcolor{black}{Security Requirements from BEoT Paradigm}}
\label{Sec:Security_Opportunities}
\textcolor{black}{This section presents the necessary requirements of BEoT paradigms through some of the key benefits of EoT protection.} Today, modern businesses use a vast, growing systems of wireless devices and data-intensive applications \cite{numan2020systematic}. As more devices are added and computing power moves closer to the device, traditional networks will not be able to maintain the level of performance required by the businesses \cite{xu2019computation}. The nature of the work accomplished by IoT devices creates a need for much faster connections between the data center and the devices \cite{tariq2019security}. \textcolor{black}{Edge computing moves computational power relatively close to the users, applications and devices where data is generated and { the actions are needed to be taken}. Approaching the data source closer can bring positive real business impacts such as better user experience, improved performance, data security, uninterrupted operation \cite{xu2019become}. In today's increasingly digital world, cybersecurity is a top concern for business, government and individuals.  As millions of devices connect, hackers find new vulnerabilities to deliver increasingly sophisticated attacks, making it much harder for systems to identify, protect and respond to these threats \cite{neshenko2019demystifying}. In addition to stealing intelligence or disrupting business activity, hackers now have more entry points allowing them to damage our physical world and pose serious security risks to utilities, factories, transportation and other critical infrastructure. Blockchain technology is one of the solution{s} to {meet} these security requirements in EoT through transparent transactions. The blockchain ledger catalogues each transaction series from end to end, enabling the reliability, synchronisation and tracking of all transactions. \cite{singh2020blockchain}.}

\subsection{Access Authentication for Edge of Things}
Smart IoT technologies are designed to make our lives simpler. Various cellular networks offer seamless connectivity for billions of things or devices. To protect the exchange of data, device manufacturers need to provide unique and reliable digital identities and ensure secure data exchange \cite{sisinni2018industrial}. Blockchain provides security against hacking, enables end-to-end encryption of the data they share \cite{savelyev2018copyright}. 

\textcolor{black}{Some previous studies employed blockchain technology to protect EoT applications like smart grids, smart transport, smart medical devices, smart cities, etc.} Some researchers focused on efficient authentication and data sharing between different platforms \cite{chaudhary2019best}. The work in \cite{guo2019blockchain} introduces a method for improving distributed, trusted authentication services on blockchains and the EoT. Byzantine error tolerance consensus algorithm was proposed to develop a blockchain for data storage and authentication. Edge computing was applied to a blockchain by providing two edge nodes, a resolution edge node, and a cache node. Resolution edge nodes provide name resolution, and the caching node aims to provide edge authentication using smart contracts and helps to improve the hit ratio. The asymmetric cryptography model was proposed to address security challenges between terminals and nodes. The experimental results show the algorithm's efficiency in terms of effective communication and computing costs, while the proposed model outperforms existing models by reducing the delay rate of 6\%-12\% and increasing the hit rate 8\%-14\%. However, the proposed model can be enhanced by reducing latency while transferring large data packets to the destination. 

In a real-time environment, achieving minimal latency with high security is a challenging task. In \cite{wang2019blockchain}, the authors proposed an authenticated blockchain model with an effective key agreement protocol for the smart grid edge-computing systems. Experimental results promise security improvements with minimal latency for smart grid growth. The proposed model focused more on providing better security, although the computing cost can be minimized by maintaining the ES cache nodes. In a similar wok in \cite{mahmood2018pairing}, the authors proposed a secure key agreement protocol using blockchain for smart grid edge computing systems. The main requirement of this proposed model is that the smart meter sometimes fails to check the authenticity of the electrical power control, and therefore, the authentication process is not achieved at a better rate. Transferring goods safely from source to destination using supply chains requires high bandwidth, which can be achieved with 5G enabled EoT. In \cite{jangirala2019designing}, the authors proposed blockchain-based authentication technology integrated with the RFID supply chain system in 5G enabled EoT for efficient computing and communication costs. The proposed authenticated model works on cryptographic hash and bitwise XOR rotation.  Initially, the authors considered N blocks, and each block has the privilege of a reader tag. The reader tag must prove its identity by transferring the authentication message to the supply chain. The supply chain validates the received message and ensures acknowledgment. Experimental results achieve a higher security rate with effective communication costs compared to existing models. In addition, the proposed model can be extended further to focus on a real-time problem.

The rapid growth of vehicular edge computing (VEC) in smart transport has intensified the implementations on traffic systems. Accessibility of communication channels, authentication of privacy and trust management in automobiles have made VEC highly prevalent. In \cite{liu2020blockchain}, the authors proposed a VEC blockchain model based on trackable map directions using dynamic route hash chain. This model's vision is to build a decentralized, secure system with low communication overhead. Moreover, the proposed model does not achieve better latency and communication overhead for a 256-bit data message, thus inhibiting its usefulness in VEC. Another interesting work to provide authentication in electric vehicles integrated with cloud infrastructure and edge computing  \cite{liu2018blockchain}, the authors proposed blockchain-based data coins and energy coins on a decentralized network. During this process, blockchain technology enhances authenticated data processing and security mechanisms for energy transmission. However, the proposed model does not specify how data manipulation, identification are carried out, thus limiting its use in the VEC. Another application of blockchain for efficient data sharing in VEC can be found in \cite{kang2018block}, where the data can't be shared without proper authorization. The vehicular blockchain model uses smart contracts to accomplish effective and reliable information storage on roadside units (RSUs) and information sharing within automobiles. The reputation-based access control technique is used to make sure the transmission of reliable information between vehicles. The experimental results for the detection of abnormal vehicles at a trust threshold of 0.35 is 100\% for the proposed model, while the other existing model is only 50\%.  
\subsection{Data Privacy for Edge of Things}
\textcolor{black}{Data privacy is one of the key requirements that protect data from malicious access. A number of data privacy mechanisms available include encryption, decryption, perturbation-based, and blockchain. Data is securely transmitted in the blockchain by maintaining timestamps and hash functions. Shared information is distributed across multiple sites using a distributed ledger \cite{alkadi2020deep,he2017efficient}.}
In \cite{8874972}, researchers propose a privacy-preserving method by assigning tasks to the edge nodes using smart contracts, in which each block keeps the assigned task information. All edge nodes connected to a decentralized network and the information is distributed using alias function in the blockchain. Edge nodes need to perform the assigned task and calculate time and energy consumption. Experimentation performed on a variety of privacy methods to prevent the storage of block information from multiple data mining threats. Moreover, the proposed model achieved a satisfactory privacy rate; however, it can be improved by focusing more on optimizing energy consumption at edge nodes. The successful development of the smart grid depends on the transformation of secure communication technologies, as the smart grid offers multiple options for collecting electrical data \cite{he2016privacy}. However, smart grid applications face challenges like energy security and privacy protection. In \cite{gai2019permissioned}, the authors proposed preserving BEoT 's privacy for smart grid applications. In this process, electricity consumption can easily be traced without disclosing end-user information to identify inappropriate energy-using behaviors by raising alarms using blockchain. Few supernodes are deployed in the blockchain responsible for resource allocations that validate the edge nodes. Here edge nodes are considered smart meters, power sensors. These edge nodes distribute the energy to the end-user, which reduces the burden on the central system and helps to improve the computing process. The edge node is validated using the covert channel authorization scheme, and the access control scheme. Validation is designed to ensure that a 51\% attack ensures that the majority of participants are good. Optimal allocation of energy resources will be made through a smart contract, covering three elements, including energy consumption, latency, and security of communication. The work in \cite{wan2019blockchain}, introduces a distributed IIoT model for smart factory using blockchain. In order to ensure proper privacy, the authors introduced the bell-la padula (BLP) approach, which is integrated with the biba model \cite{lin2012access}. Experimental findings show that the proposed model provides enhanced security and privacy features. However, the proposed model failed to achieve proper resource allocation strategies, thereby reducing its usefulness. In \cite{pyoung2019blockchain}, the authors proposed an innovative blockchain model for edge-based IoT architectures called LiTichain with multiple blocks, each with a finite lifespan. The block will be removed from the chain if the life of the transaction expires. LiTichain is created by merging two different graphs. One graph represents the life of the transaction, and the second graph represents the formation of a block in the chain. As the number of transactions increases the height of the chain, the authors have introduced a K-height block method to restrict the height of the chain. The experimental results are obtained by taking New York taxi IoT data, which transmits sensed data to the ES. The ES will collect and process the data. The proposed blockchain model is used on ES to ensure sensed IoT data privacy.

\subsection{Attack Detection for Edge of Things}
\textcolor{black}{Due to the {proliferation} in IoT sensing technology, attackers can attack and steal sensitive and vital data. Some applications, such as smart grids, smart cities, supply chains, healthcare, etc., are often used to generate sensitive IoT data and there is a high probability of attacking these data.  Cyber attackers are exploring different vulnerabilities to exploit highly sophisticated attacks, making it extremely difficult for systems to identify, protect, and respond to such attacks. The attack detection system is one of the requirement to monitor the communication system and to protect against malicious attacks \cite{singh2020blocked}.}
Recently, the authors in \cite{medhane2020blockchain} introduced a new blockchain architecture by integrating edge, cloud and SDN to achieve confidentiality and strengthen the security mechanism by preventing IoT devices from various types of malicious attacks. During this process, IoT sensors from different locations capture the data and transfer the captured data to the edge-cloud for pre-processing. The SDN-enabled blockchain process allows dynamic network traffic management and detects malicious attacks. In the cloud layer, most attacks are identified and eliminated, which reduces storage space and increases the rate of latency while reaching the edge layer; therefore, the rate of attack is drastically reduced, increasing the performance in terms of throughput and delay. Experimental results obtained by deploying 100 nodes in the \textcolor{black}{3000 m $\times$ 3000 m search area}, taking into account energy consumption, packet delivery ratio, throughput and delay as performance metrices. The results promise that the proposed security model will consume less energy and improve the transfer of packets with better throughput and delay. In addition, the proposed model does not produce any results for the detection of attacks, thus limiting its utility in the blockchain model. Another blockchain framework can be observed in \cite{pu2020rpeds}, where the economic denial of sustainability (EDoS) are prevented from malicious attackers. Secret sharing scheme (SSS) is introduced to provide security whenever the ES fails.  Sometimes, whenever the ES is down due to some malicious attack by an attacker, the proposed model uses a binary search mechanism to identify and locate the afflicted ES. The results reveal that the proposed model uses 128-bit ciphertext data, 256-bit Diffie-Hellman key, requires 0.004 ms for encryption, and 0.0039 ms for decryption. The total computational time taken by the proposed model for uploading and accessing data is 14.1199 ms. In addition to computational performance, the proposed model achieves a better attack prevention rate. Event-driven messages (EDMs) in vehicle networks will be generated during the occurrence of accidents, road slipping. EDMs consists of photos, videos, etc. and faces several challenges, such as security and latency, during the transmission of these messages. The work in \cite{nkenyereye2020secure} introduces a reliable blockchain platform with 5G-enabled vehicle edge computing to transfer EDMs to end-users by optimizing communication costs. During this process, EDMs are transferred to nearby ES in order to reduce the response time. Blockchain technology is used at the edge nodes to track messages and protect messages from a variety of attacks. The results show that the proposed model protects EDMs from different types of attacks, like impersonation attacks, DDoS attacks, Masquerade attacks, and reduces communication overheads. Another interesting work related to VEC \cite{dai2020deep}, the authors integrated deep reinforcement learning (DRL) and blockchain into vehicle networks aim of providing smart and reliable caching content. During this process, initially the proposed blockchain model ensures a decentralized data caching system in which the vehicles perform data caching and maintains an authorized blockchain at the nearby fixed base stations. Later DRL is used to develop an optimal data caching model by considering mobility as one of the metrics. Finally, to enhance the process of block verification, the authors used the new Block Verifier Selection Method, proof-of-utility (PoU).

\subsection{Trust Management for Edge of Things}
\textcolor{black}{Due to the rapid growth of technical advances, large {amounts} of data are gathered from edge nodes or IoT devices, but data protection, trust management and privacy are very important requirements on ES, particularly when the collected data is malicious and can cause serious problems.} The work in \cite{zhaofeng2019blockchain} introduces a blockchain-based, trusted data management system (BlockTDM) in edge computing. In  this process, the blockchain model is designed to ensure mutual authentication, smart contract and flexible consensus. The proposed BlockTDM ensures the privacy of data through the provision of a multi-channel data segment. The data is encrypted using user-defined encryption techniques just before the transaction is stored in the blockchain. Decryption and transaction of data in a secure blockchain is carried out using hyperledger as a smart contract. Another exciting blockchain application can be found in \cite{xiao2020reinforcement}, which preserves MEC from fake service record threats and malicious edge threats. The authors proposed an RL algorithm to decrease computational latency, optimize energy consumption, and reduce the resource allocation time of the edge devices. The experimental results show that the authors used the blockchain model on Ethereum, PoW protocol is used to promtly build service records in the blockchain. The proposed model reduces the malicious attack rate by 66.4\%, optimizes energy consumption by 10.5\%, and reduces latency by 67.4\%. In \cite{cui2019decentralized}, the authors introduced a decentralized, trustworthy blockchain model in edge computing. The domain name server sends the user request to the appropriate ES, which reduces the propagation delay. To achieve trustworthiness and security, all participants involved in transactions must share their block information and transaction details. Participants contributing to the network will earn blockchain tokens. The experimental results show that the proposed model provides 12.54\% optimal latency rate compared to the other existing model. Another interesting work in \cite{chuang2020tides} suggests a trust-aware IoT data economic system (TIDES) to provide safe, precise, and intelligent IoT data trading systems for the end-users. In the first step, the trustworthiness mechanism obviates malicious distributors to ensure secure transmission of data. In the second phase, the game-theory based pricing method facilitates win-win transactions where suppliers get better quality information at a reasonable rate and distributors get huge returns. In the third phase, if the candidate has accidentally made a transaction to a malicious distributor, the payment of the transaction will be reflected automatically. In the final phase, TIDES uses an MEC model to reduce latency and overhead storage. The experimental results show that TIDES accomplishes better results in terms of trading time, reduced latency, better security and communication costs.

\begin{table*}[]
\centering
\caption{Review of security requirements of the BEoT paradigm.}
\label{tab:Security}
\resizebox{\textwidth}{!}{%
\begin{tabular}{|c|c|c|p{7.75cm}|p{4.75cm}|}
\hline
\textbf{\begin{tabular}[c]{@{}c@{}}Security \\ services\end{tabular}} &
  \textbf{Ref.} &
  \textbf{\begin{tabular}[c]{@{}c@{}}Application \\ Domain\end{tabular}} &
  \multicolumn{1}{c|}{\textbf{Contributions}} &
  \multicolumn{1}{c|}{\textbf{Challenges}} \\ \hline
\multirow{7}{*}{\begin{tabular}[c]{@{}c@{}}\\ \\ \\ \\ \\ \\ Access \\ authentication\end{tabular}} &
  \multirow{2}{*}{\cite{guo2019blockchain}} &
  \multirow{2}{*}{\makecell{IoT \\system}} &
  Edge nodes provide name resolution, and the caching node provides edge authentication using smart contracts &
  Poor latency and delay while transferring large data packets \\ \cline{2-5} 
 &
  \multirow{4}{*}{\cite{wang2019blockchain}} &
  \multirow{4}{*}{smart grid} &
  1. The key agreement protocol ensures secure communication between the end user and the ES &
  Results limited to authentication did not show computation cost results
\\
&&& 2. Smart contract ensures secure transaction, identity verification, recording of the public key &
   \\ \cline{2-5} 
 &
  \multirow{3}{*}{\cite{mahmood2018pairing}} &
  \multirow{3}{*}{smart grid} &
  The key agreement protocol enables smart meters to acquire reliable power services from distribution control through a single private key &
  Smart meter sometimes fails to check the authenticity of the electrical power control \\ \cline{2-5} 
 &
  \multirow{3}{*}{\cite{jangirala2019designing}} &
  \multirow{3}{*}{\makecell{service \\system}} &
  Cryptographic hash and bitwise XOR rotation are used for authentication.The reader tag must prove its identity by transferring the authentication message to the supply chain &
  Proposed method not investigated on a real-time issue \\ \cline{2-5} 
 &
  \multirow{2}{*}{\cite{liu2020blockchain}} &
  \multirow{2}{*}{\makecell{vehicular \\network}} &
  Trackable map directions using dynamic route hash chain and to build a decentralized, secure system with low communication overhead &
  The model does not achieve better latency and communication overhead for a 256-bit message \\ \cline{2-5} 
 &
  \multirow{2}{*}{\cite{liu2018blockchain}} &
  \multirow{2}{*}{\makecell{vehicular \\network}} &
  Authenticated data processing and security mechanisms for energy transmission &
  Proposed model does not specify how data manipulation, identification are carried out \\ \cline{2-5} 
 &
  \multirow{2}{*}{\cite{kang2018block}} &
  \multirow{2}{*}{\makecell{vehicular \\network}} &
  Smart contracts are used to accomplish effective and reliable information storage on roadside units &
  Resource allocation at edge nodes is excluded \\ \hline
\multirow{4}{*}{\begin{tabular}[c]{@{}c@{}}\\ \\ \\ \\ Data \\ privacy\end{tabular}} &
  \multirow{2}{*}{\cite{8874972}} &
  \multirow{2}{*}{\makecell{IoT \\ network}} &
  Edge nodes connected to a decentralized network and the info is distributed using alias function in the blockchain &
  Proposed model can optimize the energy consumption at edge nodes \\ \cline{2-5} 
 &
  \multirow{2}{*}{\cite{gai2019permissioned}} &
  \multirow{2}{*}{smart grid} &
  Few supernodes are deployed in the blockchain responsible for resource allocations that validate the edge nodes &
  The model does not specify traffic load and resource allocation as the network size increases \\ \cline{2-5} 
 &
  \cite{wan2019blockchain} &
  IIoT &
  BLP approach integrated with the Biba model \cite{lin2012access} to ensure data privacy &
  Proposed model failed to achieve proper resource allocation strategies \\ \cline{2-5} 
 &
  \multirow{4}{*}{\cite{pyoung2019blockchain}} &
  \multirow{4}{*}{\makecell{IoT \\ network}} &
  LiTichain blockchain model is created by merging two different graphs. One graph represents the life of the transaction, and the second graph represents the formation of a block in the chain &
  Poor latency and delay while the number of transactions increases \\ \hline
  \multirow{10}{*}{\makecell{Attack \\detection}} &
  \multirow{2}{*}{\cite{medhane2020blockchain}} &
  \multirow{2}{*}{IIoT} &
  The SDN-enabled blockchain process allows dynamic network traffic management and detects  malicious attacks &
  Proposed model does not produce any results for the detection of attacks \\ \cline{2-5} 
 &
  \multirow{2}{*}{\cite{pu2020rpeds}} &
  \multirow{2}{*}{\makecell{service \\ system}} &
  EDoS are prevented from malicious attackers, SSS is to provide security whenever the ES fails &
  Proposed method not investigated on a real-time issue \\ \cline{2-5} 
 &
  \multirow{3}{*}{\cite{nkenyereye2020secure}} &
  \multirow{3}{*}{\makecell{vehicular \\ network}} &
  Reliable blockchain platform with 5G-enabled vehicle edge computing to transfer EDMs to end-users by optimizing communication costs &
  System design does not focus on anonymity \\ \cline{2-5} 
 &
  \multirow{2}{*}{\cite{dai2020deep}} &
  \multirow{2}{*}{\makecell{vehicular \\ network}} &
  DRL-blockchain   aims to provide smart and reliable caching content on vehicle networks &
  \multirow{2}{*}{Proposed model uses tiny dataset} \\ \hline
  
\multirow{4}{*}{\begin{tabular}[c]{@{}c@{}}\\ \\ \\ \\ \\ \\ \\ \\ Trust \\ management\end{tabular}} &
  \multirow{5}{*}{\cite{zhaofeng2019blockchain}} &
  \multirow{5}{*}{\makecell{IoT \\ network}} &
  1. BlockTDM ensures the privacy of data through the provision of a multi-channel data-segment & The system design does not reduce communication overhead \\
  &&& 2. Data encryption is carried out using user-defined encryption techniques and decryption is done by hyperledger as a smart contract & \\ \cline{2-5} 
 &
  \multirow{5}{*}{\cite{xiao2020reinforcement}} &
  \multirow{5}{*}{\makecell{Mobile \\ computing}} &
  1. RL algorithm to decrease computational latency time, optimize energy consumption, and reduce  the resource allocation time of the edge devices & The proposed model does not specify resource allocation as the size of the network increases \\ &&& 2. PoW protocol is used to build service records in the blockchain quickly & 
   \\ \cline{2-5} 
 &
  \multirow{4}{*}{\cite{cui2019decentralized}} &
  \multirow{4}{*}{\makecell{Storage \\ network}} &
  1. The domain name server sends the user's request to the ES to reduce the propagation delay & Results did not show computation cost results \\
  &&& 2. To achieve trustworthiness and security, all participants share their block information and transaction details & \\ \cline{2-5} 
 &
  \multirow{4}{*}{\cite{chuang2020tides}} &
  \multirow{4}{*}{\makecell{IoT \\network}} &
  1. Game-theory based pricing method facilitates  win-win transactions & \multirow{4}{*}{Not effective for large real-time data} \\
  &&& 2. TIDES uses MEC model to reduce latency and overhead storage & \\ \hline
\end{tabular}%
}
\end{table*}

\subsection{Summary}
\textcolor{black}{In this section, we examined the security requirements of the BEoT paradigm and benefits of blockchain in providing essential security services to EoT, such as access authentication, data privacy, attack detection, and trust management}. In today's world of advancements in Internet, wireless technology, and data-intensive applications, we have seen significant technological changes in data communication and networking applications. Edge computing is a trending technology designed to improve latency and increase computational performance. As millions of devices connect, hackers find new vulnerabilities to exploit sensitive and confidential data.  Blockchain technology can remove all these security problems through transparent transactions. We summarize security opportunities from the BEoT Paradigm in Table~\ref{tab:Security}.

\section{Research Challenges and Future Directions}
\label{Sec:Challenges_Future-Directions}
This section presents the key research challenges and future directions related to the BEoT paradigm.
\subsection{Research Challenges} 
BEoT has the  potential to spot its avenues in almost all kinds of digital applications. BEoT paradigm is an integration of three giant technologies, namely blockchain, Edge computing, and IoT. It offers significant benefits combating many issues in the performance of deploying each other separately. Therefore the issues concerned with these technologies should be addressed.Some of the challenges of the BEoT paradigm are discussed here.
\subsubsection{Security in blockchain}
Blockchain is a shared, secured, immutable, decentralized, and valid ledger, which records and tracks the transactions done on digital resources without the necessity of  centralized authority in various domains such as smart healthcare and smart cities. It enables two users to exchange and communicate in a peer-to-peer network where the distributed decisions are taken by considering the majority vote instead of a single centralized administration. Blockchain has demonstrated its ability in many applications which involves a centralized ledger. Some of the promising applications of blockchain are monitoring the network and providing security services which includes privacy, confidentiality, and integrity. Despite several potential applications of blockchain in various domains, it still has many open-ended challenges. The various security, privacy, and scalability challenges of blockchain are cryptokey management, data privacy in chain management, transaction linkage, and compliance with regulations with respect to data privacy. Several research works have been carried out on the privacy of the users in various digital scenarios. Blockchain technology was developed to deploy the Bitcoin cryptocurrency and resolves the double spending issue. The solution for this problem is bitcoin in which all the transactions are made public in the ledger. Any node can track and watch the transactions which are spent. But the problem is the complete anonymity is not guaranteed in bitcoin \cite{ bernabe2019privacy}. In the distributed EoT, ES are distributed at the network edge, making the ES vulnerable to security attacks. The conventional cryptographic techniques are difficult to be accommodated in ES as they are resource-constrained.  This challenge in EoT opens up a need for a secured lightweight authentication where the ES can authenticate the end devices at a faster pace. Furthermore, the edge server needs a trust management mechanism to ensure reliable trust computation between end nodes and various ES as these servers cannot carry trust among other servers. 

\subsubsection{Standardization}
Blockchain was originated as an infrastructure to provide solutions for the digital cash problem. Also, it allows payment across borders irrespective of the constraints in the geographical area over the Internet. Whereas it takes many days to transfer funds between various banks located in different countries using the conventional banking system. This open nature of blockchain technology makes it further expanded to address many commercial problems in several financial applications, Industrial sectors, IoT, supply chain, etc. But the speed and extent of implementation of blockchain technology are obstructed by its interoperability challenges. These challenges are not only due to the representation of various digital tokens and cryptocurrencies but also the vital differences in the behavior of transaction management. This makes blockchain difficult to combine with other conventional enterprise systems and interoperate. This eventually creates issues for the regulatory acceptance of blockchains, thereby raising a need for standardization.

\subsubsection{Resource management in BEoT}
The decentralized blockchain framework empowers the robustness and scalability of the system by utilizing the optimal resources from all the nodes, thereby reducing the latency in the data processing as well as making the resource-restricted IoT platform a robust resource utilization framework. On the other hand, IoT encompasses the devices with restricted bandwidth, whereas blockchain consumes more bandwidth. BEoT, a distributed platform where the heterogeneous data (distributed in different areas) from heterogeneous nodes are accumulated and processed in a distributed environment. Therefore, the various distributed resources like data centres for storage, robust machines for complex computation, interoperable middleware services, user details can be managed effectively. The distributed ledger at every node in blockchain alleviates the need for a centralized server by storing the device credentials and transactions. As the number of IoT devices is increasing, the number of transactions in the ES also increases rapidly. Therefore higher processing capability and computational network resources are required to ensure the increased processing capability and low latency responses.

The storage overhead incurred due to this massive processing requirement of blockchain in processing real-time data streaming, can be reduced by segregating the metadata required from the original data stream and minimizing the contents to be stored on the blocks. Though lightweight blockchain serves this purpose, its scalability will decrease with an increase in the number of network nodes. Therefore, sidechains can be used with blockchain as a control layer \cite{moniruzzaman2020blockchain}. The major problem in the integration of IoT and blockchain is storage constraint, i.e., a combination of resource-constrained IoT with high resource-consuming blockchains. Resource-constrained edge devices with limited computing resources are not efficient in handling numerous transactions when integrated with huge resource consuming blockchain frameworks. The decentralized blockchain framework trust mechanisms will ensure trustworthiness in data from edge nodes \cite{ xu2020novel}.

\begin{enumerate}
\item {\textbf{Scalability:}}
Scalability is a more significant issue in the data storage of cloud-centric IoT devices. As the blockchain grows with the number of users or transactions, it is difficult for the IoT devices to store the ledgers as it increases in size. Furthermore, IoT devices range from low-power to high-end servers. So, depending on the resource capability of the IoT device, designing a device-specific blockchain is a trending challenge. This includes the security algorithms, efficient mining process, and appropriate metadata segregation from the ledger. 
\item
{\textbf{Intelligence:} 
The BEoT paradigm offers various services like secure and privacy-preserving data sharing with low latency response for various industrial applications. It still lacks intelligence in processing and prediction the future behaviour of the applications. For instance, earlier prediction of disease in smart healthcare based on the information accumulated in the edge, demand response prediction and EVs lifetime prediction in smart grids, traffic data prediction in intelligent transportation systems, resource demand prediction in smart cities and predictive maintenance of home appliances requires intelligent agents to enable predictive analytics in almost all industrial applications of BEoT.}
\end{enumerate}

The resource management issues concerned with various applications of Blockchain, IoT, and edge computing individually as well as in combination is presented in Table~\ref{tab:mytab3}. It is evident that resource management directly impacts the acceptability, scalability, robustness, interoperability, load provisioning, long-term sustainability, faster data processing (with low-latency response), and task offloading. Henceforth, to design a scalable and secured BEoT environment hosting of the resources, metadata segregation in blocks, faster (higher bandwidth) and controlled access to the resources must be enforced.


\subsection{Future Directions}
This section presents the various solutions and future directions in BEoT with AI and  5G networks. 
\subsubsection{Solutions to Research Challenges}
Various research solutions in the literature had addressed the challenges in the BEoT environment.
\begin{itemize}
    \item \textbf{{Security Frameworks:}}
    The work in \cite{salman2018security} discusses various security problems and services provided by blockchains. One of the advantages of blockchain is its pseudo-user anonymity feature, but there is a threat that the user information can be exposed to hackers. The users are not anonymous, in which blockchain-based access control lists(ACL) associate the ACL with the users directly. The same problem occurs in blockchain-based provenance and key management. The anonymity problem of blockchain is solved by bitcoin in which the public key of the user is their identification. A trust enabled blockchain framework called trustchain for privacy preserving transactions in EoT was proposed in \cite{jayasinghe2019trustchain}. Trustchain, lightweight permissioned blockchain (consent violation) ensures privacy preservation among its prosumers and avoids unprecedented delays in distributed networks.  Furthermore, as the edge computing moves computing resources closer to the end users, the privacy of data should be ensured in distributed computing environments.
\item {\textbf{Standardization bodies:}}
    An initiative for standardization has been started on blockchain through a professional committee of International Organization for Standardization (ISO), World Wide Web Consortium (W3C) International Telecommunication Union (ITU), Institute of Electrical and Electronics Engineering (IEEE) and Internet Engineering Task Force (IETF). ISO develops standards for nomenclature, terminology, ontology and architecture, privacy, identity, security, interoperability, smart contracts, and Governance. The W3C is a standardization organization produced web standards and they have started a community group for blockchain to develop standards for message format, guidelines for storage in private and public blockchain, torrent, and side chain. The ITU has created a focus group on distributed ledger technology (DLT) to recognize and examine the services and applications of DLT, to create guidelines and practices for the implementation of these services and applications in the global market. IEEE has created a project for blockchain; namely standardization initiative for the blockchain framework in IoT. IETF defines suite for Internet protocol for interoperability standards and network communication for blockchain technologies \cite{ gramoli2018blockchain}.
    \item {\textbf{Scalable Architectures:}} 
    Skyline Queries are added benefits in optimal dataset query processing.  These queries will help to retrieve the results from an optimal-related set instead of searching the whole dataset. This, in turn, reduces the data processing time and removes the overhead in storing larger datasets\cite{ ferrag2018blockchain}. The resource management framework using blockchain proposed in \cite{ xu2017intelligent} alleviates the tremendous amount of energy consumed by processing the explosive data accumulation at the cloud data centers. Though edge computing offers low latency response, edge nodes have a limited capacity, which makes it difficult when the user demand increases \cite{xiong2018mobile}. This leads to constrained access to IoT devices and undetermined network latency. The virtual resources are hosted on the edge nodes with blockchain for managing these transactions \cite{ samaniego2016hosting}. Access control to the devices is provided by blockchain through the management hub \cite{novo2018scalable} on the edge of the sensor networks. Furthermore, these IoT devices are more vulnerable to various security attacks because of their resource constraints. So a permission blockchain with Edge computing, namely EdgeChain, was developed in \cite{pan2018edgechain}. 
\end{itemize}

\subsubsection{Other Future Directions and Enabling Technologies}
Though multiple issues were solved in literature, the BEoT should be compatible and upgraded itself for known and unknown future problems. Therefore these challenges should be addressed before the full-fledged adoption of  BEoT.  Some of the future directions in the BEoT environment that evolves as the technology advances are usability, cybersecurity, memory management, Access control for resources and users, real-time data stream delivery,  and predictions on future trends and patterns\cite{alamri2019blockchain}.

\textit{Enhancing Blockchain Performance for the Betterment of BEoT:}
Though blockchain is secure, the blockchain security issues discussed {in terms of} challenges and the vulnerabilities imposed by ES and IoT devices will have a greater impact on the future BEoT framework. Furthermore, the ability of blockchain to store entire transaction data may create storage burden leading to scalability issues. This ability of blockchain consumes more energy, network bandwidth and thereby reducing the throughput. As the IoT devices will be increasing rapidly, the scalability and storage issues will significantly impact the performance. Therefore a lightweight consensus mechanism should be incorporated for blockchain mining processes by segregating the appropriate data from the ledger and storing it in side chains \cite{moniruzzaman2020blockchain}. Also, data inconsistency due to proliferation of nodes in the lightweight blockchain will remain. Although skyline queries provide effective data processing, data privacy still remains a baffling issue \cite{ferrag2018blockchain}. A lightweight blockchain framework with efficient data processing and service validation, alleviating the scalability issue will be a research direction for future BEoT. Bigdata processing systems will be an essential candidate to handle the enormous data accumulations. Also, the bigdata processing will provide effective processing of data in resource constrain{ed} ES to ensure low latency response.

\textit{Integration of AI and BEoT:}
As BEoT lacks intelligence required for predictive analysis in smart applications, AI will provide a more sophisticated solution. AI aids the machines to mimic the natural intelligence possessed by humans \cite{nguyen_enabling_2020}. These intelligent agents (AI incorporated machine) were programmed to simulate the cognitive behavior of the human and their brain, which is specifically utilized for solving problems in real-time through experience (learning) \cite{ russell2002artificial}. AI have bought remarkable automation in many domains utilizing computational schemes. This led to the revolution of smart computing systems. Some of them are robots in military applications (mission-critical activities), health care assisted robot in monitoring patients in the absence of surveillance, automated transportation (anonymous operated autonomous vehicles), gaming applications, content delivery network (routing),  marketing (making all predictions on single search), chatbots (online agent, a virtual assistant to handle customers) in finance, agricultural robots, rovers in space research, and social networking (predictive analysis).

The notable subsets of AI are ML and deep learning (DL), the most significant technological advancements. Indeed, there are numerous technical challenges in the BEoT paradigm that can be assisted by the AI’s smartness. Privacy preservation is the prominent issue in the mission-critical application of BEoT standard.  The study in \cite{nawaz2019edge} have suggested the integration of AI at edge nodes aiming at the higher level of the privacy to the users with fewer security vulnerabilities by alleviating the use of third parties to mine the data at the edge. The proposed model suggests the storage of processed data in blocks instead of storing the raw data and retrieving it back for processing later. Edge AI allows the end devices to control all the mining processes through Ethereum smart contracts (where the data owners can update their policies involved in data processing and sharing) directly, which enables to reduce considerable network bandwidth. Furthermore, they indicate that data owners at the edge can process their data using neural network algorithms and make valuable predictions as the immutable nature of blockchain records all the transaction involved in mining.

Labelling of datasets in AI empowers the model. It provides better performance, but the traditional crowdsourcing (generating labels from the human input) is centralized and have various issues {such as} data privacy and delayed processing. Furthermore, training the DL models from the local edge server {requires larger storage and computing power}, which is not feasible with resource-constrained edge devices. The training of models can be shifted to the cloud and can be retrieved for the processing, which may lead to data privacy issues.  Therefore, Edgence, a framework proposed in \cite{9089179}  provides decentralized crowdsourcing and decentralized AI training with blockchains for secured and privacy-preserving data transactions in decentralized IoT applications. Thus, blockchain ensures the privacy of the data allowing the data owners to negotiate their data usage by third parties and the robust AI algorithms processing at the edge, thereby drastically reducing the network latency and the load on a block in a blockchain transaction. 
The survey in \cite{yang2019integrated} have explored the broader perspectives of blockchain and edge computing applications. One of the more extensive areas addressed is the contribution of AI in blockchain-based edge computing applications and its benefits. With evidence from the literature, they have suggested that AI will tremendously impact the performance of BEoT, especially in resource management, automatic generation of smart contracts, prediction of faults, and scalable off-chain computation at the network edge. 

In-edge AI framework in \cite{wang2019edge} was proposed for the effective utilization of the interaction among the mobile edge nodes to train the AI models with a reduced computational load. They have employed the DL techniques like DRL to attain optimal caching at edge nodes with minimal computation and federated learning for managing the resources effectively.  Furthermore, they suggest that the blockchain framework can be integrated with this environment, but load distribution among heterogeneous edge nodes remains unexplored and will be a research direction.

\textit{Cognitive Edge:}
Cognitive edge is an interesting idea in the edge computing space that utilizes the cognition from AI algorithms for processing at the edge effectively. A blockchain-based optimal knowledge paid sharing for AI-enabled edge nodes was introduced in \cite{ lin2019making}. For effective knowledge aggregation with reduced computational load, knowledge management chain and knowledge trading subchains are deployed. ML algorithms are used for knowledge extraction. AI-enabled edge nodes are priced for knowledge sharing due to their imposed selfishness. The blockchain consortium based on smart contracts uses knowledge coins for knowledge trading among Edge AI nodes. This ensures tamper-resistant (knowledge coins are stored in knowledge managers), decentralized and very fair trading. The green blockchain Proof-of-Trading combines the PoS and PoW for reduced resource consumption.  Thus the optimal pricing will encourage smooth knowledge trading among buyers and sellers in AI edge nodes. 

On the other hand, sharing the economy-related services in smart environment, leverages computational load as well as incurs the security burden. Sharing economy related services and smart contracts \textcolor{black}{have a tremendous} impact on the massive BEoT crowd. Blockchain and AI-based cognitive edge framework \textcolor{black}{was designed} in \cite{rahman2019blockchain} for facilitating the sharing economy-related service in a smart city. The proposed model uses DL algorithms for extracting the meaningful and significant event information from the IoT data. The framework uses cognitive edge nodes for storing and processing the immutable ledgers of blockchain and off-chain (includes the transactions involving IIoT and mobile edge devices). The model integrates the cognitive processing at the edge for sharing economy-related services. AI is used for data processing and extracting shareable economy-related data from heterogeneous IoT devices, blockchain and social network media. 

Furthermore, as the number of IoT devices in the smart environment increases, the traffic generation will increase tremendously leading to higher back-haul data rates. Intelligent caching is performed to overcome this, by storing popular contents at different locations of the network. But still, this has a more significant challenge in decision making based on future content popularity. This can be resolved with DL and reinforcement learning for significant decision making on cache content prediction \cite{khan2020edge}.
Therefore, when AI is integrated with the BEoT environment, it reduces the computational power, increases data processing rate and scalability, increases the decision making in load management and intelligent caching, and predicts any system attacks in advance. The major security concerns and resource optimization with the use of blockchain for EoT can be alleviated with the power of CE. All these can be achieved without human intervention but through the intelligent agents mimicking the human cognition. The greatest challenge in this adoption is to focus on the effective training of the AI models in heterogeneous environments. Because the machine is trained through learning from previous experience, but training or learning in a dynamically changing environment is challenging and can impact the overall performance. One of the most common solutions suggested in the literature is retrieving information from all the heterogeneous nodes for training the model, but its effectiveness is still unexplored and paves a way for future research.

\textit{BEoT in 5G:}
The emerging 5G network is 20 times faster than the current 4G LTE used in almost all cellular networks\cite{ chaer2019blockchain}. It can carry a huge payload in a shorter duration by its variety of spectrum bands. The scalable BEoT environment requires higher computational resources with low latency response. Also, BEoT applications demand peer to peer communication for its transactions, 5G has the more suitable  capabilities for hosting the BEoT. Both technologies drive each other forward. As 5G sorts out the communication constraints in blockchain and blockchain ensures the privacy concerns. 5G infrastructure crowdsourcing using smart contracts, 5G infrastructure sharing (i.e., national roaming and spectrum sharing) without a third party, the most challenging international sharing, network slicing to accommodate multiple users with seamless interaction, massive machine communication and low-latency ultra-reliable communications. The massive adoption of 5G with blockchain has particular hindrance such as managing throughput with scalability, transforming an enormous number of contracts in 5G into smart contracts, need for regulatory compliance, privacy, cloud infrastructure costs, and a trusted registration system\cite{ chaer2019blockchain}. Usually, the more powerful 5G cellular network with edge computing is integrated with AI for effective mining of big data accumulated in the edge nodes as ineffective mining may degrade the system performance. And blockchain cryptocurrencies namely, bitcoin and litecoins can be utilized for the privacy-preserving virtual transaction \cite{chang2020guest}. Also, the major cost incurred in handling the cloud infrastructures can be reduced by using Edge intelligence (edge computing with distributed AI). The edge enabled 5G-blockchain-based infrastructure for scheduling distributed heterogeneous edge resources was proposed in \cite{zhang2019edge}.

The intelligent transportation framework in \cite{xie2019blockchain} was developed to address the reliable security scheme requirement of the vehicular ad-hoc network (VANET) with enhanced vehicle trust management in traffic monitoring. 5G-VANET with SDN is used to ensure the utmost reliability and global traffic network control. The immutable ledger blockchain with centralized authentication is used to secure the traffic system from malicious vehicles (the vehicle that violates the traffic rules and regulation) by creating hazardous traffic in the network. Blockchain stores the vehicle details, traffic tag details of its travel path and vehicular messages in the blocks. Also, blockchain is utilized to offer trust management by computing the trust value through the integration of PoW and PoS. The simulation results guarantee the better privacy preserved trust management for IoT vehicular networks. 
The use of blockchain in 5G MEC unlocks the new business value, brings new value shift, and captures this value in the telecommunication industry. The low-latency communications of 5G are attained through edge computing. The privacy concerns in heterogeneous MEC are solved by blockchain. The blockchain is constructed to enhance user privacy as well as privacy of network topologies (attained using accommodative bloom   filter without revealing the topology privacy by maintaining the routing   consensus) \cite{ yang2020distributed}. Their integration is fulcrum behind the excellent results in 5G BEoT supply chain management \cite{jangirala2019designing} and unmanned vehicular systems\cite{9113786}.

\begin{table*}[!ht]
\caption{Challenges, Future Directions and Benefits of BEoT.}
\label{tab:mytab3}
\resizebox{\textwidth}{!}{
\begin{tabular}{|c|p{4cm}|p{2.2 cm}|p{5.3 cm}|p{4.3 cm}|}
\hline
\textbf{Ref.} & \textbf{Challenges} & \textbf{Application} & \textbf{Description} & \textbf{Benefits} \\ \hline
 \multirow{2}{*}{\cite{moniruzzaman2020blockchain}} & Massive real-time data streaming and storage overhead in the lightweight blockchain & Smart home and surveillance systems & Segregation of metadata through lightweight blockchain to  scale down processing time and edge computing to reduce latency & Acceptability, long-term sustainability and  scalability \\ \hline
 \multirow{2}{*}{\cite{ ferrag2018blockchain}} & Inefficient mining process when the users are increased and skyline query processing in  blockchain & Any smart systems & Energy-efficient consensus protocol design for mining the application-specific data to be stored on the ledger & Scalability, robustness and faster data   processing \\ \hline
 \multirow{2}{*}{\cite{ xu2017intelligent}} & Massive explosion of data will consume more energy and the cost will be higher & Smart grid system & Decentralized  blockchain-based resource management with embedded reinforcement learning for request migration in smart contracts & Cost minimization in energy consumption in   power grids \\ \hline
 \multirow{2}{*}{\cite{xiong2018mobile}} & Load distribution when the networks scale up devices and safety loading & Cloud-centric IoT  & Provisioning of  virtual resources and permissioned blockchain for access control in edge nodes & Low-latency response and secured transaction \\ \hline
 \multirow{2}{*}{\cite{ samaniego2016hosting}} & Edge computing resource allocation & Mobile blockchain & Number of entries in the block must be optimal, an economic model with optimal resource utilization & Optimal resource utilization, low-latency and scalability \\ \hline
 \textbf{Ref.} & \textbf{Future direction} & \textbf{Application} & \textbf{Description} & \textbf{Benefits} \\ \hline
 \multirow{2}{*}{\cite{novo2018scalable}} & Global storage of different resource   access control details  into the blockchain & IoT devices in smart environment & Decentralized   blockchain where the access control policy of the entire system is stored in   a single blockchain & Scalability \\ \hline
 \multirow{2}{*}{\cite{pan2018edgechain}} & On-demand resource provisioning to   heterogeneous IoT devices & Smart systems & 
Permitted blockchain to link IoT devices and the resources on the edge nodes  and credit (internal coins)-based resource management & Scalability, secure auditing and data logging \\ \hline
 \multirow{2}{*}{\cite{nawaz2019edge}} & BEoT mining at edge nodes by alleviating raw content in the block & Edge AI for smart health care & Edge AI is used for local decision making at edge nodes. & Reduced network  resource consumption at the edge   
\\ \hline
 \multirow{2}{*}{\cite{yang2019integrated}} & BEoT resource  management and reinforced security management policies & AI’s neural   networks in blockchain & ML and DL algorithms improve the efficacy of BEoT paradigm with reduced energy in distributed computation & Scalable mining at the edge, low computation overhead and vulnerability analysis \\
 \hline
 \multirow{2}{*}{\cite{ wang2019edge }} & MEC task   offloading, resource management and load distribution when integrated with   blockchain & AI learning models in blockchain & DRL to attain optimal caching at edge nodes with minimal computation and federated learning for resource management & Cognitive computing, effective   task offloading and optimal edge caching \\ \hline
 \multirow{2}{*}{\cite{ lin2019making} }& Knowledge trading for AI-based BEoT environment & Knowledge gaining via. ML and DL & Knowledge chain with sidechains is used for decentralized, tamper-resistant, confidential and fair pricing of AI-enabled edge nodes & Aggregated resource management and fair knowledge pricing\\ \hline
 \multirow{2}{*}{\cite{   chaer2019blockchain}} & Transforming an enormous number of contracts   in 5G into smart contracts & Blockchain for 5G & System must ensure scalable transactions in handling numerous smart  contracts with low cost and secured authentication mechanisms & Scalability and standard regulatory compliance for blockchain\\ \hline
 \multirow{2}{*}{\cite{ xie2019blockchain}} & Trust management in SDN enabled   5G vehicular adhoc network & 5G intelligent transportation system & Blockchain assures privacy concerns in 5G  with 5G with vehicle privacy and secured traffic monitoring & Trust management and malicious node detection  \\ \hline
 \multirow{2}{*}{\cite{ zhang2019edge}} & Secure edge services under more   complex industrial networks & Blockchain with 5G in IIoT & A DRL algorithm is used for edge resource management (cross-domain sharing) in 5G   beyond IIoT applications & Cross-domain resource sharing and scheduling, tamper-resistant resource management \\ \hline
 \multirow{2}{*}{\cite{9113786}} & Extracting untapped value in 5G  MEC & 5G BEoT in supply chains and UAV & Describes how blockchain absorbs the value created by the 5G MEC in the telecommunication   value chain & Automated new business value creation permanent, verifiable and transparent transactions \\ \hline
 \multirow{2}{*}{\cite{ yang2020distributed}} & Privacy protection of MEC & Trust management in diversified MEC & Blockchain for ensuring user privacy and network privacy in multi-server collaboration & Trusted routing in collaborative   network and privacy in network topology \\ \hline
\end{tabular}%
}
\end{table*}
 The summary of challenges, future directions, its application and benefits of BEoT is described in Table \ref{tab:mytab3}.

\section{Conclusion} 
\label{Sec:Conclusion}
In this article, we have conducted an extensive survey of \textcolor{black}{the use of blockchain in EoT networks and \textcolor{black}{associated} applications}. We have first introduced an overview of the blockchain and EoT and discussed the main motivations behind \textcolor{black}{the use of blockchain in EoT networks}. Furthermore, we have also provided a generic BEoT architecture where IoT, edge computing, blockchain and succeeding applications and security services have been analyzed. \textcolor{black}{Subsequently}, we have paid attention to the review of the BEoT adoption in a number of important industrial applications, including smart transportation, smart city, smart healthcare, smart home, and smart grid. The security benefits of the BEoT paradigm have been discussed, with some key services such as access authentication, data privacy preservation, attack detection, and trust management. Finally, we have outlined some research challenges and pointed out open research directions toward BEoT-5G networks. We believe that this article will instigate exemplary approaches on BEoT research for future applications and services. 

\balance

\bibliography{Ref}
\bibliographystyle{IEEEtran}

\end{document}